\documentclass[letterpaper,twocolumn,10pt]{article}
\usepackage{usenix2019_v3}

\usepackage{array}
\usepackage{booktabs}
\usepackage[small,bf]{caption}
\usepackage{url}
\usepackage{textcomp}
\usepackage{xspace}
\usepackage{titling}
\usepackage{listings}
\usepackage{flushend}
\usepackage{graphicx}
\usepackage{pifont}

\usepackage[compact]{titlesec}

\usepackage[scaled=0.8]{beramono} % Replace courier
\usepackage[T1]{fontenc}

\usepackage{microtype}

\usepackage{soul} % for diff markup

\usepackage{hyperref}
\hypersetup{% squlech lacheck
    unicode=false,          % non-Latin characters in Acrobat’s bookmarks
    pdftoolbar=true,        % show Acrobat’s toolbar?
    pdfmenubar=true,        % show Acrobat’s menu?
    pdffitwindow=false,     % window fit to page when opened
    pdfstartview={FitH},    % fits the width of the page to the window
    pdftitle={},  % title
    pdfauthor={},     % author
    pdfsubject={},   % subject of the document
    pdfcreator={},   % creator of the document
    pdfproducer={}, % producer of the document
    pdfkeywords={}, % list of keywords
    pdfnewwindow=false,      % links in new window
    colorlinks=true,       % false: boxed links; true: colored links
    linkcolor=magenta,          % color of internal links (change box color with linkbordercolor)
    citecolor=magenta,        % color of links to bibliography
    filecolor=magenta,      % color of file links
    urlcolor=black           % color of external links
}

\usepackage{enumitem}
\setitemize{noitemsep,topsep=0pt,parsep=0pt,partopsep=0pt}
\setenumerate{noitemsep,topsep=0pt,parsep=0pt,partopsep=0pt}
\setdescription{itemsep=0pt,topsep=0pt,parsep=0pt,partopsep=0pt,itemindent=0em,leftmargin=0.5em}

\newcommand{\us}{\textmu{}s\xspace}
\newcommand{\faster}{\texttt{FASTER}\xspace}
\newcommand{\hlog}{\texttt{HybridLog}\xspace}

\iffalse % change to false to generate without diff markup.
\newcommand{\old}[2][]{{\color{red}\st{#2}\textsuperscript{#1}}}
\newcommand{\new}[2][]{{\color{blue} #2\textsuperscript{#1}}}
\else
\newcommand{\old}[2][]{}
\newcommand{\new}[2][]{#2}
\fi

\newcommand{\one}{\ding{172}\xspace}
\newcommand{\two}{\ding{173}\xspace}
\newcommand{\three}{\ding{174}\xspace}

\begin{document}

%don't want date printed
\date{\vspace{-4eX}}

%make title bold and 14 pt font (Latex default is non-bold, 16 pt)
\setlength{\droptitle}{-4em}
\title{\Large \bf Achieving High Throughput and Elasticity in a
Larger-than-Memory Store}

\author{Chinmay Kulkarni\\
University of Utah\\
chinmayk@cs.utah.edu
\and
Badrish Chandramouli\\
Microsoft Research\\
badrishc@microsoft.com
\and
Ryan Stutsman\\
University of Utah\\
stutsman@cs.utah.edu}

\maketitle

% Use the following at camera-ready time to suppress page numbers.
% Comment it out when you first submit the paper for review.
% \thispagestyle{empty}
%\pagenumbering{gobble}

\begin{abstract}

Millions of sensors, mobile applications and machines now
generate billions of events.
%
% These events are processed and
% aggregated in services in the cloud to gain insights and drive application
% logic.
%
Specialized many-core key-value stores (KVSs) can
ingest and index these events at high rates (over 100 Mops/s on one machine)
if events are
generated on the same machine; however,
%
% in practice, these events need
% to be aggregated from a wide and distributed set of data sources. Hence, fast
% indexing schemes alone only solve part of the problem.
%
to be practical and
cost-effective they must ingest
events over the network and scale across cloud resources elastically.
% , provisioning and reconfiguring over inexpensive generic
% cloud resources as workload demands change.

We present Shadowfax, a new distributed KVS based on \faster{}, that
transparently spans DRAM, SSDs, and cloud blob storage while serving
130~Mops/s/VM over commodity Azure VMs using conventional Linux TCP. Beyond
high single-VM performance, Shadowfax uses a unique approach to distributed
reconfiguration that avoids any server-side key ownership checks
or cross-core coordination both during normal operation and migration.
Hence,
Shadowfax can shift load in 17~s to improve system throughput by
10~Mops/s
with little disruption. Compared to the state-of-the-art, it has 8$\times{}$ better throughput
  (than Seastar+memcached) and \new{avoids costly I/O to move cold data during migration.}
\new{On 12~machines, Shadowfax retains its high throughput to
perform 930~Mops/s},
which, to the best of our knowledge, is the highest
reported throughput for a distributed KVS used for
large-scale data ingestion and indexing.

\end{abstract}

\section{INTRODUCTION}

%{\color{red} Notes from call with Badrish:
%Move 3 goals to end of intro, rewrite into short narrative form instead of
%list. In ideas list: expand partitioned point to be clearer. Combine async
%clients with global cut; bring in word sessions. Kick hw accel to end where
%we say session batching works well with cloud hw accel offerings for low-cost
%networking/dispatching. Finally, say sessions play key rol ein parallel data
%migration.}

\begin{figure}[t]
\centering
\includegraphics[width=0.93\columnwidth]{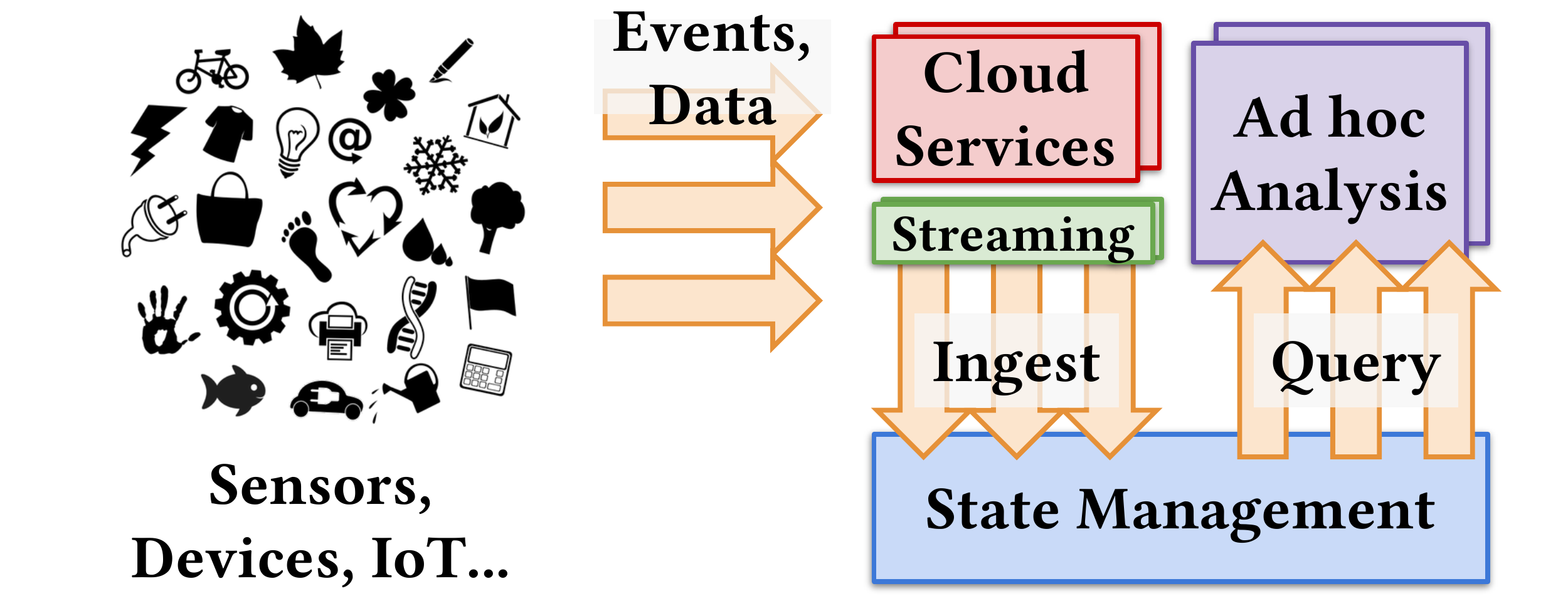}
\caption{
A typical data processing pipeline.}
% Services receive and
% process raw events. A state management system ingests processed
% events and serves offline queries against them.}
\label{fig:pipeline}
\end{figure}

% XXX This could be stronger if it gave a time window and had a citation.
Millions of sensors, mobile applications, users, and machines
continuously generate billions of events that are
are processed by streaming
engines~\cite{spark-streaming, trill} and ingested and aggregated
by state management systems (Figure~\ref{fig:pipeline}).
Real-time queries are issued against this ingested data to train and
update models for prediction, to analyze user behavior, or to generate
device crash reports, etc.
Hence, these state management systems are a focal point for massive numbers of
events and queries over aggregated information about them.

This has led to specialized KVSs that can
ingest and index these events at high rates (100 million operations (Mops) per second (s) per machine)
using many-core hardware~\cite{faster,anna}.
They are efficient if events are generated on the same machine as
the KVS, but, in practice, events must be aggregated from wide
and distributed sets of data sources.
Hence, fast indexing only solves part of the problem.
To be practical and cost-effective, a complete system for aggregating
events must ingest them over the network, must scale across machines
and cores, and must be elastic (by provisioning and reconfiguring over
inexpensive cloud resources as workloads change).

Existing KVSs with similar
performance~\cite{mica, flexnic, floem, kvdirect} rely on application-specific
hardware acceleration, making them impossible to deploy on today's cloud
platforms.
These systems also only store data in DRAM and do not scale across
machines; adding support to do so without hurting normal-case performance
is not straightforward.
For example, many of them statically partition records across cores to
eliminate cross-core synchronization.
This optimizes normal-case performance, but it makes concurrent
operations like migration and scale out impossible; transferring records
and ownership between machines and cores requires a stop-the-world approach
due to these systems' lack of fine-grained synchronization.

Achieving this level of performance while fulfilling all of these
requirements on commodity cloud platforms requires solving two key challenges
simultaneously.
First, workloads change over time and cloud VMs fail, so systems must tolerate
failure and reconfiguration.
Doing this without hurting normal-case performance at 100~Mops/s is hard, since
even a single extra server-side cache miss to check key ownership or
reconfiguration status would cut throughput by tens of millions of operations
per second.
Second, the high CPU cost of processing incoming network
packets easily dominates in these workloads,
especially since, historically, cloud networking stacks have not been designed
for high data rates and high efficiency.
We show this is changing; by careful design of each server's data path, cloud
applications can exploit transparent hardware acceleration and offloading
offered by cloud providers to process more than 100~Mops/s per cloud virtual
machine (VM).

We present \emph{Shadowfax}, a distributed KVS built over \faster, our 
high-performance open-source single-node KVS\footnote{\faster{} is available at \url{https://github.com/microsoft/FASTER}.}. Shadowfax transparently
spans DRAM, SSDs, and cloud storage while serving 130~Mops/s/VM on
commodity Azure VMs~\cite{azure} with conventional Linux TCP.
Beyond high per-VM performance, its unique approach to
distributed reconfiguration avoids any server-side key ownership checks
and any cross-core coordination during
normal operation and data migration both in its indexing and network interactions.
Hence,
it shifts load in 17~s to improve cluster throughput by
10~Mops/s
with little disruption.
Compared to the state-of-the-art, it has 8$\times{}$ better throughput (than
Seastar+memcached~\cite{seastar}) while avoiding I/O to move cold data
during migration (compared to Rocksteady~\cite{rocksteady}).

In this paper, we describe and evaluate three key pieces of Shadowfax that
eliminate coordination throughout the client and server side by eliminating
cross-request and cross-core coordination:
\begin{description}
\item[Low-cost Coordination via Global Cuts:]
% All of the major components of Shadowfax including indexing, request
% dispatching, durability, checkpointing/recovery, and migration center
% around a key mechanism: asynchronous \emph{global cuts}~\cite{faster,cpr,scalog}.
%
In contrast to totally-ordered or stop-the-world approaches used by most
systems, cores in Shadowfax avoid stalling to synchronize with one another, even when
triggering complex operations like scale-out, which require
defining clear before/after points in time among concurrent operations.
Instead, each core participating in these operations -- both at clients and
servers -- independently decides a point in an \emph{asynchronous global
cut} that defines a boundary between operation sequences in these complex operations.
In this paper, we extend asynchronous cuts from cores within one process~\cite{faster,cpr} to servers
and clients in a cluster, and we show how they coordinate server
and client threads (through partitioned sessions) by detailing
their role in Shadowfax's low-coordination data migration and
reconfiguration protocol.

%for complex operations like checkpointing, recovery, and
%reconfiguration, Shadowfax makes pervasive use of asynchronous \emph{global
%cuts}.
%
%In this paper, we mainly focus on how global cuts help coordinate both server
%and client threads through partitioned sessions by focusing in detail on
%their role in Shadowfax's low-coordination data migration and
%reconfiguration protocol.
%%
%To maintain high throughput when moving ownership of records between two
%server VMs, Shadowfax takes a \emph{global cut} across
%server threads, and propagates it independently to client threads.
%%
%This allows the index to be shared under normal operation
%while avoiding serial bottlenecks during scale out.

\item[End-to-end Asynchronous Clients:]
All requests from a client on one machine to Shadowfax are asynchronous with
respect to one another all the way throughout Shadowfax's client- and
server-side network submission/completion paths and servers' indexing and
(SSD and cloud storage) I/O paths.
This avoids all client- and server-side stalls due to head-of-line
blocking, ensuring that clients can always continue to generate requests and
servers can always continue to process them.
In turn, clients naturally batch requests, improving server-side high
throughput especially under high load.
This batching also suits hardware accelerated network offloads available in
cloud platforms today further lowering CPU load and improving throughput.
Hence, despite batching, requests complete in less than 40~\us to 1.3~ms at
more than 120~Mops/s/VM, depending on which transport and hardware
acceleration is chosen.
%
% Our experiments show that this combination works well for both linux TCP
% and two-sided RDMA.

\item[Partitioned Sessions, Shared Data:]
Asynchronous requests eliminate blocking {\em between requests} within a client, but
maintaining high throughput also requires minimizing coordination
costs {\em between cores} at clients and servers.
Instead of partitioning data among cores to avoid synchronization on record
accesses~\cite{hstore,voltdb,mica,seastar}, Shadowfax partitions network
sessions across cores; its lock-free hash index and log-structured record heap
are shared among all cores.
This risks contention when some records are hot and frequently
mutated, but this is more than offset by the fact that no software-level
inter-core request forwarding or routing is needed within server VMs.

\end{description}

The rest of the paper is organized as follows. We provide background on the \faster{}
key-value store and its use of epochs within a machine (\S\ref{sec:fasterkv}). Next,
we overview Shadowfax's design, including partitioned client
sessions with global cuts and how they enable
reconfiguration~(\S\ref{sec:design}). We then provide details on our parallel
non-blocking migration and scale-out techniques~(\S\ref{sec:scale-out}).
%
\iffalse
In the remainder of this paper, we describe how the key synchronization
mechanisms at the core of \faster{}'s design (\S\ref{sec:epochs}) naturally led
to Shadowfax's sessions that extend global cuts over the network
(\S\ref{sec:sessions}). We describe how this enables Shadowfax to perform the
same over the network as with a local \faster{}
instance~(\S\ref{sec:eval:clients}), and we describe how they enable
reconfiguration~(\S\ref{sec:ownership}) and parallel data
migration~(\S\ref{sec:scale-out}). We also describe how Shadowfax does this
while supporting larger-than-memory datasets that span SSD and cloud blob
storage.
\fi
%
Next, we evaluate Shadowfax in detail against other state-of-the-art shared-nothing
approaches~(\S\ref{sec:eval}), showing that by eliminating record ownership
checks and cross-core communication for routing requests it improves
per-machine throughput by 8.5$\times$ on commodity cloud VMs.
We also show it retains high throughput during migrations and scaled it
to a
cluster that ingests and indexes \new{930~Mops/s},
which, to the best of our knowledge, is the highest
reported throughput for a distributed KVS
used for large-scale data ingestion and indexing.
Finally, we cover
related work~(\S\ref{sec:related}) and conclude~(\S\ref{sec:conclude}).

\section{BACKGROUND ON \faster{}}
\label{sec:fasterkv}

Shadowfax is built over the \faster single-node KVS, which it relies on for
hash indexing and record storage.
Here, we describe some key aspects of \faster, since Shadowfax's design
integrates with it and builds on its mechanisms.
More details about \faster itself can be found elsewhere~\cite{faster,cpr}.
Specifically, Shadowfax extends \faster's asynchronous
cuts, which help avoid coordination, and its \hlog, which transparently
spans DRAM and SSD.

\begin{figure}[t]
\centering
\includegraphics[width=\columnwidth]{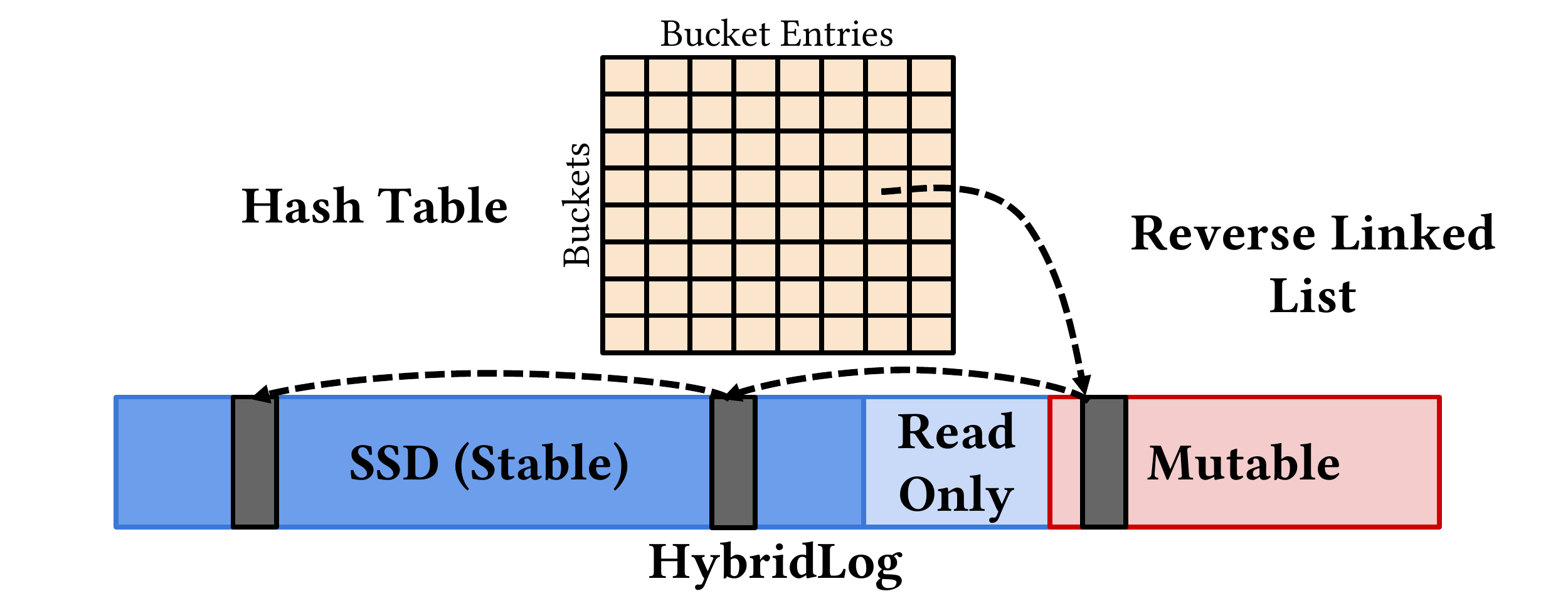}
\caption{\faster's \hlog spans memory and SSD.}
%    The portion in memory contains a mutable region that acts as a
%    cache and a read-only region.
%    \faster's hash table points to a reverse linked list of
%    records on the \hlog.}
\label{fig:hlog}
\end{figure}

In most ways, \faster works like most durable hash table libraries.
It includes a lock-free hash table divided into cacheline-sized buckets
(Figure~\ref{fig:hlog}).
Each 8~byte bucket entry contains a pointer to a record whose key hashes to
that bucket.
Each record points to another record, forming a linked list of records with
common significant key hash bits.
Each bucket entry contains additional bits from the associated records' key
hash, increasing hashing resolution and disambiguating what records the bucket
entry points to without extra cache misses and without full key comparisons.
Each record pointed to by the hash table is stored in the \hlog.

\faster clients can use it like any other library, but a common pattern is to
pin one client application thread per CPU core to eliminate scheduler overheads.
Each client thread calls read or read-modify-write operations on keys in
\faster.
\faster's cache-conscious design and lock-freedom are key in its ability to
perform more than 100~Mops/s on a single multicore machine.
%
%Shadowfax uses \faster with a similar threading scheme described later.

\subsection{\hlog Allocator}
\label{sec:hlog}

\faster allocates and stores all records in its \hlog, which spans memory and
SSD (Figure~\ref{fig:hlog}).
The \hlog combines in-place updates (for records in memory) and log-structured
organization (for records on SSD), and provides lock-free access to records.

The portion of the \hlog's address space on SSD forms the stable
region.
It contains cold records that have not been recently updated.
The portion in memory is composed of two regions: a (larger) mutable region and a
(smaller) read-only region.
Records in the mutable region can be modified in-place with appropriate
synchronization that is chosen by the application using \faster (for example, atomic
operations, locks, or validation).
This region acts as a cache for recently updated records and avoids expensive
per-update allocations.

The read-only region mostly contains records that are being asynchronously
written to SSD.
These records cannot be updated in place since they must remain stable during I/O.
The read-only region represents records that are becoming cold, and it acts as
a second-chance cache.
\faster uses a read-copy-update to modify records in this region: the updated
record version is appended to the mutable region, and the hash table is updated to
point to it.
This helps provide good cache hit rates without fine-grained metadata.

Each record entry in \faster's hash table points to a reverse linked list of
records on the \hlog, allowing it to maintain a compact hash table for
\emph{larger-than-memory} datasets that span storage media.
Note, that a consequence of this is that hash table lookups in \faster may need
to traverse chains of records that span from memory onto SSD.
Section~\ref{sec:design:indirection} describes how Shadowfax extends \hlog so
that it also spans shared cloud storage and how this accelerates the completion
of scale out and data migration.

\subsection{Asynchronous Cuts}
\label{sec:epochs}

Lock-freedom makes \faster fast, but it creates challenges for synchronization
and memory safety.
Updated versions of records may be installed in its hash table, even as old
versions of that record are still being read by other threads.
This is a common problem in all lock-free, RCU-like schemes~\cite{rcu}.
To solve this, \faster uses an epoch-based memory-protection scheme~\cite{epochs}.
All threads calling into \faster are registered with an epoch manager that
tracks when threads begin and end access to \faster's internal structures.
When a page is evicted to SSD, the epoch-based scheme ensures that the memory
is not reused while any thread could still be accessing it.
The full details of this scheme are beyond the scope of this paper.

Critically, this epoch-based scheme also plays a key role in coordinating
information across threads lazily without inducing stalls.
During complex, process-wide events (such as page eviction and checkpointing), threads lazily
coordinate by registering callback actions that are eventually executed once each
thread synchronizes some local state with an updated
process-global value.
The same mechanism can also be used to trigger a function only once all threads
are guaranteed to have updated their local state from some process-global state.
In effect, this allows trigger actions that are guaranteed to
take effect only after all threads agree on and have each locally
observed some transition in process-global state.
This can be used to create a process-wide \emph{asynchronous cut}, where
events such as process state transitions are realized asynchronously and lazily
over a set of independent thread-local state transitions.

\begin{figure}[t]
\centering
\includegraphics[width=\columnwidth]{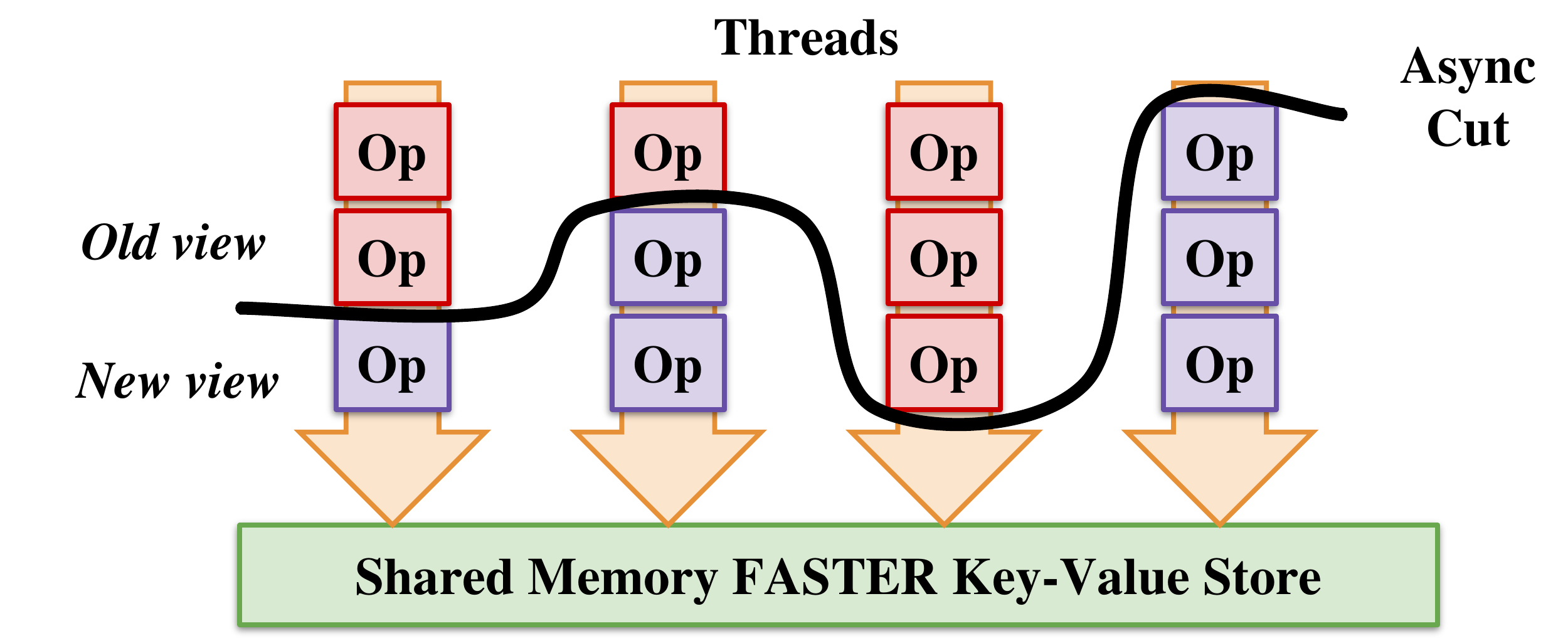}
\caption{Asynchronous cuts in \faster.}
%    View changes of shared state in \faster take place over
%    an asynchronous cut using epochs. Process-global state is updated first; when
%    every thread has observed the update a post-change function is triggered.}
\label{fig:cut}
\end{figure}

For instance, consider the read-offset address that demarcates read-only records from mutable
records on the \hlog (Section~\ref{sec:hlog}). When this address is updated, each thread may
notice the update at different points in time, depending on when they refresh their epoch.
Eventually, when all threads have observed the update, the records between the old and new
read-offsets have become read-only, and a function is triggered to write the pages to disk. Using the same
mechanism, addresses for which threads do not yet agree on the mutability status can be handled efficiently.
Figure~\ref{fig:cut} shows this process in action.

\iffalse
\faster's checkpointing protocol serves as a great example of this~\cite{cpr}.
%
It first moves all threads from a system checkpoint version $v$ to a
new checkpoint version $v+1$, and then it captures and persists records in
version $v$.
%
As threads process operations, they stamp each new record version they create
with their local checkpoint version number.
%
The boundary between the two checkpoint versions $v$ and $v+1$ forms a
cut across all of the operations of all of the threads (Figure~\ref{fig:cut}).
%
The protocol first increments a process-global variable that contains the
systems' checkpoint version number, and it registers an epoch action
that checkpoints version $v$ when all threads
have observed this new checkpoint version $v+1$.
%
Each thread periodically (or when it encounters specific situations) checks for
epoch actions, which triggers a refresh of each thread's local checkpoint version number.
%
Once each thread has refreshed its local copy of the global checkpoint
version number to $v+1$ all operations before the cut are guaranteed to have
completed.
%
Hence, it is safe to checkpoint version $v$ of the database, which is triggered
by the registered epoch action.
\fi

\faster{}'s epoch protection works within a single shared memory process on one machine. Section~\ref{design:ownership-tx} shows
how Shadowfax extends the notion of cuts to apply globally \emph{across machines} -- with the assistance of
client threads -- to safely move ownership of records between servers while preserving throughput.

\section{SHADOWFAX DESIGN}
\label{sec:design}

Shadowfax is a distributed key-value store.
Each server in the system stores records inside an instance of
\faster, and clients issue requests for these records over the network.
These requests can be of three types: \emph{reads} that return a
record's value, \emph{upserts} that blindly update a record's value, and
\emph{read-modify-writes} that first read a record's value and then
update a particular field within it.
Within a server, records are allocated on \faster's \hlog, whose stable
region is extended by Shadowfax to
also span a shared remote storage tier in addition to main memory and
local SSD.

Each server runs one thread per core, and it shares its \faster instance among all threads.
Threads on remote clients directly establish a network \emph{session} with one
server thread on the machine that owns the record being accessed (\S\ref{sec:sessions}).
Sessions are the key to retaining \faster{}'s throughput over the
network:
they allow clients to issue asynchronous requests;
they batch requests to improve server-side throughput and avoid
  head-of-line blocking;
and they avoid software-level inter-core request dispatching.

Shadowfax uses hash partitioning to divide records among servers.
The set of hash ranges owned by a server at a given logical point of
time is associated with a per-server strictly increasing \emph{view
number}.
A fault-tolerant, external metadata store (e.g.\ ZooKeeper~\cite{zookeeper})
durably maintains these view numbers along with mappings from hash ranges to
servers and vice versa.
View numbers serve two key purposes in Shadowfax.
First, they help minimize the impact of record ownership checks at servers,
helping them retain \faster's performance.
Second, they allow the system to make lazy and asynchronous progress
through record ownership changes~(\S\ref{sec:ownership}).

Sessions and low-coordination global cuts via views play a key role in
Shadowfax's reconfiguration, data migration, and scale out.
Its scale-out protocol migrates hash ranges from a \emph{source} server
to a \emph{target} server and is designed to minimize migration's
impact to throughput.
The protocol uses a view change to transfer ownership of the hash range
from the source to the target along with a small set of recently accessed
records.
This allows the target to immediately start serving requests for these
records and helps maintain high throughput during scale out.
Since views are per-server, this also ensures that multiple migrations
between disjoint sets of machines can take place simultaneously.
Next,
threads on the source work in parallel to collect records from \faster
and transmit them over sessions to the target.
Similarly, threads on the target work in parallel to receive these
records and insert them into its \faster instance.
This parallel approach helps migrate records quickly, reducing the
duration of scale out's impact on throughput.
Scale out completes once all records have been moved to the target.

\subsection{Partitioned Dispatch \& Sessions}
\label{sec:dispatch}

\begin{figure}[t]
\centering
\includegraphics[width=0.9\columnwidth]{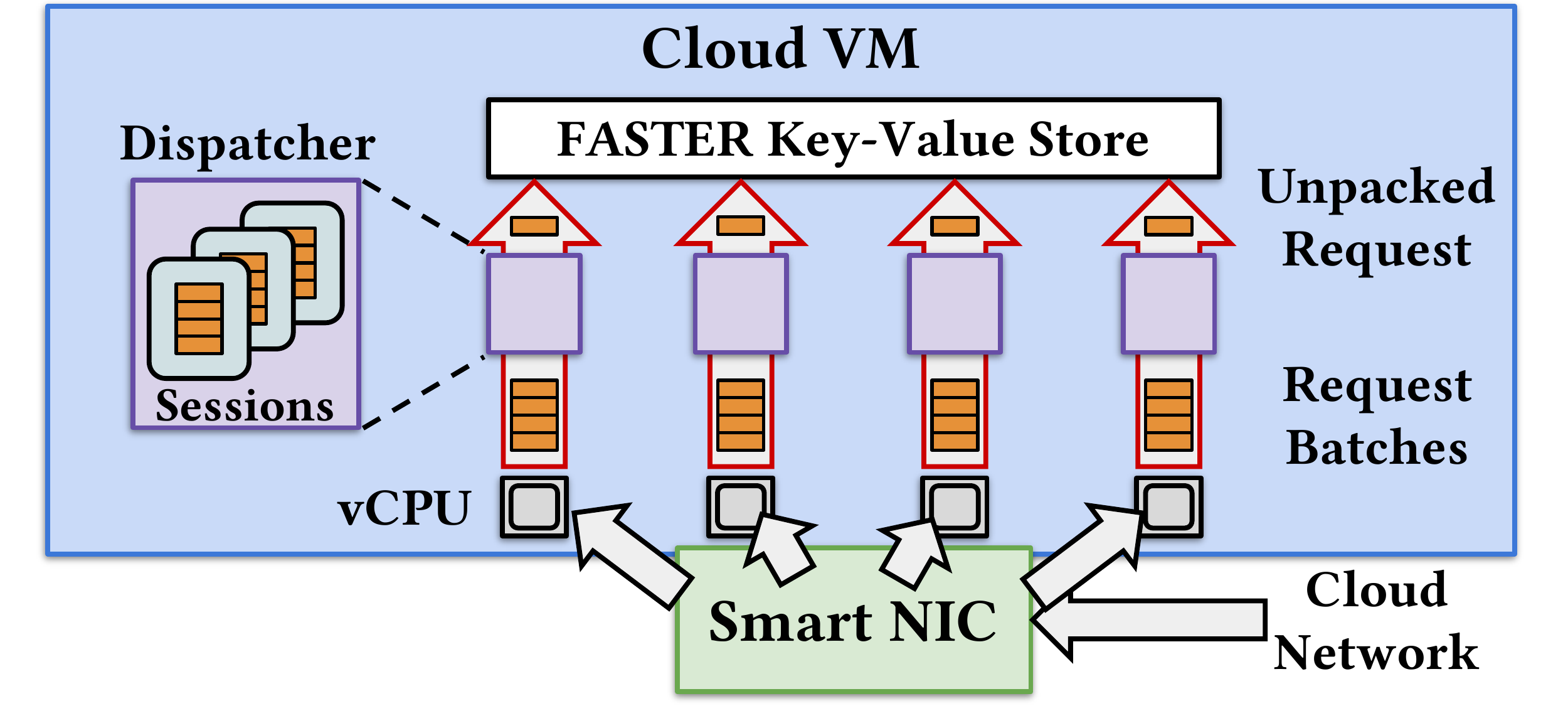}
\caption{Server threading and dispatch.}
%    Each server thread receives batches of requests from sessions
%    and processes them via a shared, per-machine \faster instance.
%    Results are returned over the network by the same
%    thread, avoiding cross-thread coordination.}
\label{fig:server}
\end{figure}

Shadowfax's network request dispatching mechanism and client library need to be
capable of saturating servers inside \faster.
One option would be to maintain a \faster instance per server thread,
partitioning records across them to avoid cache coherence costs.
However, this would create a routing problem at the server; requests
picked up from the network would need to be routed to the correct thread.
This would require cross-thread coordination, hurting throughput and
scalability.
Clients could be made responsible for routing requests to the correct
server thread, but this would require every client thread to open a
connection to every server thread and would not scale.
To avoid this, client threads could partition and shuffle requests between
themselves to directly transmit requests to the correct server thread, but this would
require cross-thread coordination at the client which would also not scale well.

Using a connectionless transport like UDP could make client-side routing
feasible without introducing cross-thread coordination~\cite{fb-memcache,mica}.
However, the system would lose its ability to perform congestion
control and flow control or tolerate packet loss, which
are basic requirements for running a
networked storage system.

Shadowfax avoids cross-thread coordination by sharing a single
instance of \faster between server threads.
\faster defers cross-core communication to hardware cache coherence on the
accessed records themselves, cleanly partitioning the rest of the system
(Figure~\ref{fig:server}).
Each server runs a pinned thread on each vCPU inside a cloud VM.
Each server thread runs a continuous loop that does two things.
First, it polls the network for new incoming connections.
Next, it polls existing connections for requests, and it unpacks these
requests, calling into \faster to handle each of them.
After requests are executed, the returned results are
transmitted back over the session they were received on.
Since \faster is shared, neither requests nor results are ever passed across
server threads.

\subsubsection{Client Sessions}
\label{sec:sessions}

Shadowfax's partitioned-dispatch/shared-data approach also extends to
clients.
Since they don't need to route requests to specific server threads, they
can reduce connection state while avoiding cross-thread coordination.

\begin{figure}[t]
\centering
\includegraphics[width=0.9\columnwidth]{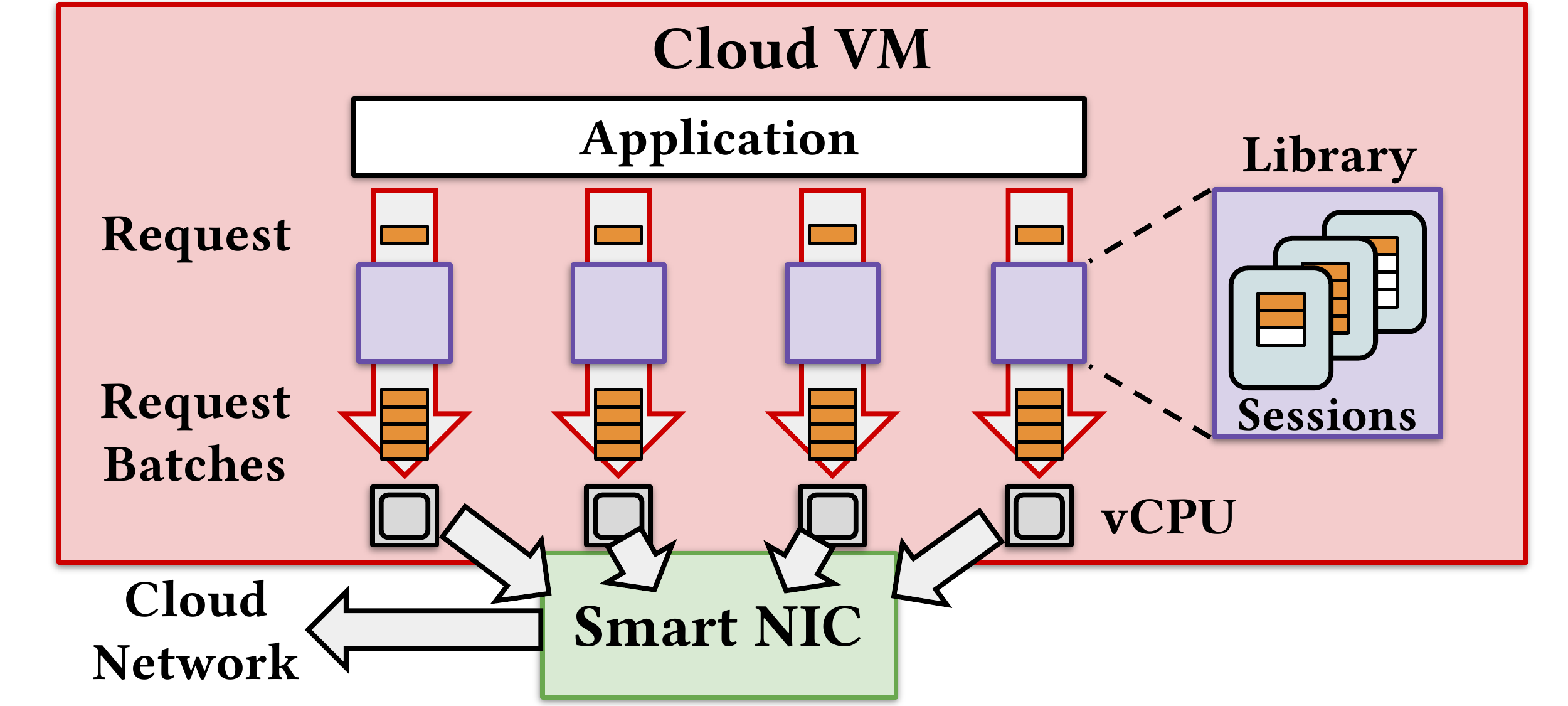}
\caption{Client threading and dispatch.}
%  Client threads partition requests into per-session transmit buffers
%  along with a callback.  Batches of asynchronous requests are kept pipelined
%  to the server, keeping both the client and the server busy.}
\label{fig:client}
\end{figure}

However, clients must also avoid stalling due to network delay in order to
saturate servers.
To do this, each client thread is pinned to a different vCPU of a cloud VM,
and it issues asynchronous requests against an instance of
Shadowfax's client library (Figure~\ref{fig:client}).
The library pipelines batches of these requests to servers.

The client library achieves this through \emph{sessions}.
When the library receives a request, it first checks if it has a
connection to the server that owns the corresponding record.
If it does not, it looks up a cached copy of ownership mappings
(periodically refreshed from the metadata store), establishes a
connection to a thread on the server that owns the record, and
associates a new \emph{session} with the connection.
Next, it buffers the request inside the session, enqueues a
completion callback for the request inside the session, and returns.
This allows the client thread to continue issuing requests without
blocking.
Once enough requests have been buffered inside a session, the
library sends them out in a batch to the server thread.
On receiving a batch of results from the server, the library
dequeues callbacks and executes them to complete the corresponding
requests.

Sessions are fully pipelined, so multiple batches of requests can be sent to a
server thread without waiting for responses.
This also means that a
client thread can continue issuing asynchronous requests into session
buffers while waiting for results.
This pipelined approach hides network delays and helps saturate servers.
It also helps keep request batch sizes small, which is good for latency.

%
% This approach to enqueuing and dequeuing callbacks assumes that server
% threads execute requests in a strict FIFO order.
%
% However, \faster executes requests for records on local SSD
% out-of-order.

\subsubsection{Exploiting Cloud Network Acceleration}

The cloud network has traditionally not been designed for high data
rates and efficiency.
The high CPU cost of processing packets over this network can easily
prevent servers and clients from retaining \faster's throughput.
However, this is beginning to change; many cloud providers are now
transparently offloading parts of their networking stack onto SmartNIC
FPGAs to reduce this cost.
Shadowfax's design interplays well with this acceleration; batched requests
avoid high per-packet overheads and its reduced connection count avoids the
performance collapse some systems experience~\cite{farm}.

Since threads do not communicate or synchronize, all CPU cycles
recovered from offloading the network stack can be used for executing
requests at the server and issuing them from the client.
This allows Shadowfax to retain \faster's high throughput using the Linux
kernel's TCP stack on cloud networks, avoiding dependence on kernel bypass
or RDMA.

\subsection{Record Ownership}
\label{sec:ownership}

To support distributed operations such as scale out and crash recovery,
Shadowfax must move ownership of records between servers at
runtime.
This creates a problem during normal operation:
a client might send out a batch of requests to a server after referring
to its cache of ownership mappings.
By the time the server receives the batch, it might have lost ownership of
some of the requested records in the batch (e.g.\ due to scale out).
Hence, the server must validate that it still owns the requested records
before it processes the batch.
This would hurt throughput if each request
was cross-checked against a set of hash ranges owned by the server.

To solve this, each set of hash ranges owned by a server
is associated with a per-server strictly-increasing \emph{view number}.
All request batches are tagged with a view number, so servers can quickly
assess whether a batch only includes requests for records that it currently owns.
When a server's set of owned ranges changes, its view number
is advanced.
Each server's latest view number is durably stored along with a list of the
hash ranges it owns in the metadata store.

When a client connects to a server, it caches a copy of the server's
latest view (a view number and its hash ranges) inside the session.
Every batch sent on that session is tagged with this number,
and clients only put requests for keys into batches that were owned by that
server in that view number.
%
%The server maintains a copy of its latest view too (periodically
%refreshed).
%
Upon receiving a batch, the server always checks its current view number
against the view number tagged on the batch.
If they match, then the server and client agree about which hash ranges are
owned by the server, ensuring the batch is safe to process without further key
or hash range checks.
If they don't match, then the client or server has out-of-date
information about which hash ranges the server owns.
Hence, the server rejects the batch and refreshes its view from the
metadata server.
Upon receiving this rejection, the client refreshes its view
from the metadata server and reissues requests from the rejected batch.

View numbers offload expensive hash range checks on each requested
key to clients, reducing server load.
For a server that owns $P$ ranges accepting $R$ requests in batches of
size $B$, views reduce the cost of checks from $O(R\log{P})$
to $O(R/B)$.
Since it is one integer comparison per batch; it also
ensures we never take a cache miss to perform ownership checks, which would be prohibitive at 100~Mops/s.
Hence, views are key in supporting dynamic movement of ownership between
servers while preserving normal case throughput.
%
% Views also provide a means for lazily propagating ownership
% changes across the system;
%
% machines not involved in an ownership change can continue operating
% normally until they refresh their ownership cache.

% RS: I started writing this here before I realize the next section basically
% addresses this. Leaving it here in case I want to steal things from it.
%
%For consistency, all server threads must agree where in the sequence of the
%server's operations an old view ends and a new view begins to take effect.
%%
%However, a server's operations are not totally ordered, and we want to avoid
%pausing server threads on view transitions to bring them into agreement about
%hash range ownership.
%%
%View changes use the global cut mechanism used in other places in \faster that
%require propagating state changes among server threads; hence, for a short
%window on the server different threads may operate in different views.

\subsubsection{Ownership Transfer}
\label{design:ownership-tx}

When ownership of a hash range needs to be transferred to or away from a
server, its ownership mappings are first atomically updated at the
metadata store.
This increments its view number and adds or removes the hash range from
its mapping.
Servers and clients observe this view change either when they refresh
their local caches of views and ownership mappings (via an epoch action) or
when they communicate with a machine that has already observed this change.

When a server involved in the transfer observes that its view has
changed, it must move into the new view.
However, this step is not straightforward; keeping with
Shadowfax's design principle, it must be achieved without stalling
server threads.
\begin{figure}[t]
\centering
\includegraphics[width=\columnwidth]{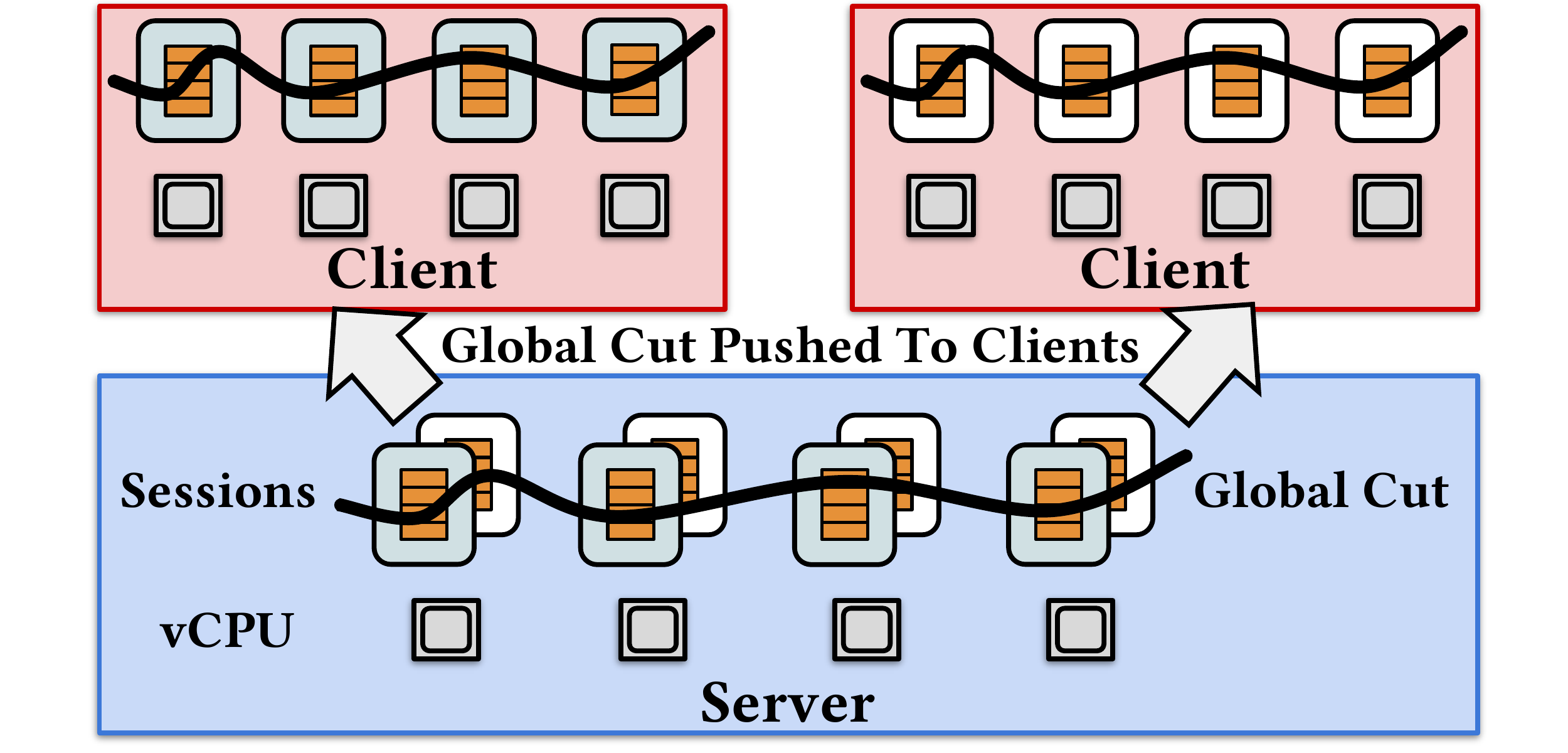}
\caption{Ownership transfer over a global cut.}
%  A view change is asynchronously propagated
%  within a server, defining a cut across server threads. Then, the server
%  extends this into a global cut covering all its connected client sessions.
%  This defines a global view boundary among all operations while
%  avoiding cross-core coordination both at servers \emph{and} clients.}
\label{fig:ownership}
\end{figure}
Within the server, this view change is propagated asynchronously across
threads via an epoch action (Figure~\ref{fig:ownership}).
Threads each mark a point in their sequence of operations, collectively creating
an async cut among all of the operations on all of the threads at the server
(\S\ref{sec:epochs}).
This cut unambiguously ensures no two servers concurrently
serve operations on an overlapping hash range.
This approach is free of synchronous coordination, helping maintain high
throughput.

The server might be connected to clients still using the old view; it must also
propagate the view change to clients in a similar way without stalling client
threads.
Sessions help Shadowfax achieve this.
%
%Recall that a session is a mechanism for sending out asynchronous
%request batches, and associates with a connection between a server
%thread and a client thread.
%
When a server thread moves into a new view, view validations on request
batches received over sessions with clients still in an older view
are rejected.
%
%These rejections (and hence the view change) are propagated over
%the sessions to all client threads connected to the server thread.
%
% This effectively pushes the async cut taken on the server out to all
% clients connected to it.
%
On receiving a rejected batch over a session, each client thread first
independently updates its thread local cache of ownership mappings and
views.
Next, the thread marks the point in the sessions' sequence of
operations after which batches were rejected by the server (since there
can be multiple such batches because of pipelining, this has to be
the earliest such point).
Collectively, these points help create an implicit async cut across
threads within a
client.
Thus, clients avoid cross-thread
coordination when observing an ownership change.
Each client connected to the server creates its async cut independently,
resulting in a cluster wide \emph{asynchronous global cut} for ownership
transfer.

Once it has observed ownership transfer, each client thread must reissue
requests that were rejected by the previous owner.
It does so by \emph{shuffling} these requests between its sessions to the
previous and new owners of the transferred hash range.
First, they are marked invalid within the previous session's
buffer.
Next, they are (re)buffered into the correct session based on
the updated ownership mappings.

\new{
These views are a form of view synchronous
communication~\cite{reliable-communication} and are similar to other
view-based approaches used for
agreement~\cite{viewstamped-replication,paxos-simple,raft}. Though, here we apply the
technique to hash range ownership rather than group membership for
replication or multicast. This approach contrasts with lease-based
approaches~\cite{leases} (e.g.\ Vertical Paxos~\cite{vertical-paxos}) that
are commonly used for this purpose~\cite{farm,ramcloud}, which depend on a synchronicity
assumption for safety and can block awaiting lease expiry in the case of slow
machines. This view-based approach sidesteps this limitation for
migrations; any agent can drive the process to completion (either a
successful migration with ownership moved to the target or a failed migration
with ownership reverted to the source), providing a form of
wait-freedom~\cite{wait-free} (aside from writes to the highly available
metadata store, which must rely on (weak) synchronicity assumptions to
ensure progress~\cite{flp}).
}

\section{SCALE-OUT AND HASH MIGRATION}
\label{sec:scale-out}
Shadowfax migrates hash ranges from a \emph{source} to a \emph{target} server
to scale out.
Migration uses global cuts to proceed in asynchronous phases that
transfer hash range ownership to the target before migrating records, as
described next.

\subsection{Migration Protocol}

Migration is implemented as a state machine on the source and target.
Both servers transition through migration phases on global cuts, created in the
same non-blocking, low-coordination way described in \S\ref{sec:epochs}.
First, each thread enters into a phase at a point in the sequence of requests
that it is processing that it chooses (a point that makes up part of the global
cut for the transition into that phase), and then it starts performing the work
of that phase.
Once all threads have entered into the phase and have completed all work
relating to it, the server transitions to the next phase.

\begin{figure}[t]
\centering
\includegraphics[width=\columnwidth]{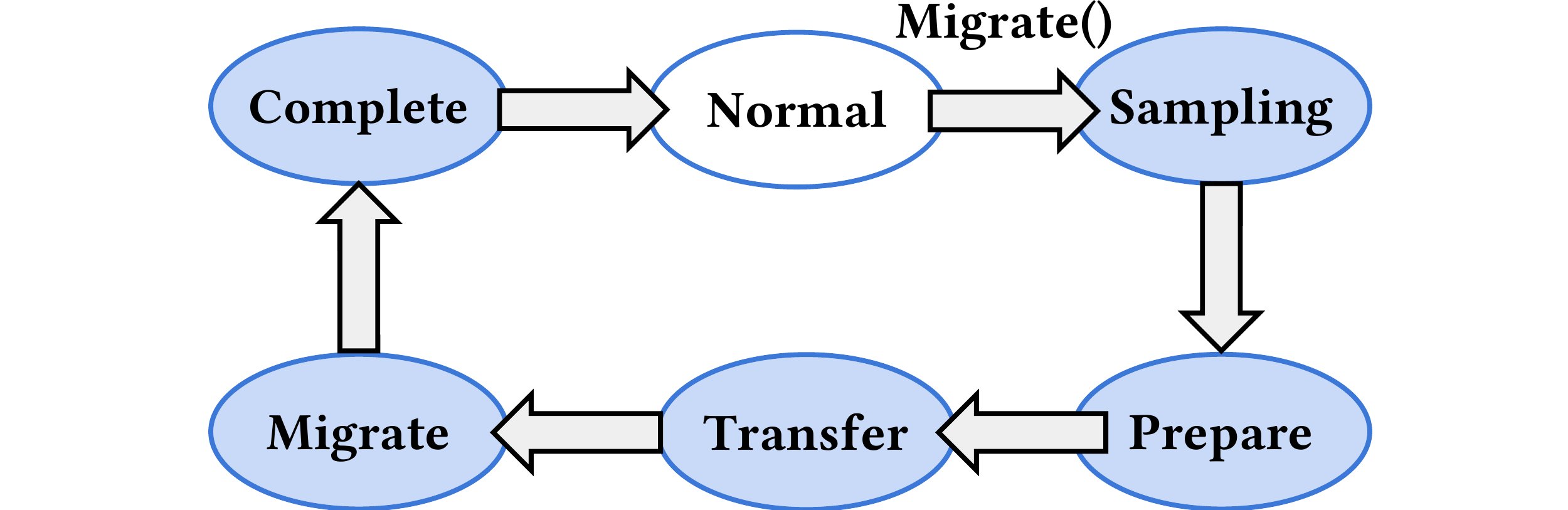}
\caption{Migration state machine on the source.}
%    This state machine is
%    responsible for
%    moving the source into the new view, for sampling and shipping hot records to the target, and for migrating all
%    records in the hash range to the target.}
\label{fig:source}
\end{figure}

Migration is driven by the source as we outline below (Figure~\ref{fig:source}):
%
%Its state machine (Figure~\ref{fig:source}) triggers the source to move into
%the new view, sampling and shipping of hot records to the target along with
%ownership transfer notification, and migration of records from the moved hash
%ranges to the target.
%
%We outline the source's phases below:% (Figure~\ref{fig:source}):

\begin{description}
\item[Sampling:]
Initiated by receiving a \texttt{Migrate()} RPC from a client, whereupon the source
\begin{enumerate}
  \item atomically remaps ownership of hash ranges from the source to the
    target, increments the source's and target's view numbers, and registers a
    dependency between the source and target (for crash recovery,
    \S\ref{sec:fault-tolerance}) within the metadata store; and
  \item begins sampling hot records by forcing all accessed
    records to be copied to the \hlog tail.
\end{enumerate}

Since the records are not yet at the target and a migration is in
progress, both the source and the target continue to temporarily operate in the
old ownership view; at this point the source is still servicing requests for
records in the migrating ranges.
To ensure that sampled records only get copied once, the source only copies
records whose address is lower than the \hlog tail address at the start of this
phase.

\item[Prepare:]
Initiated after all source threads have completed the Sampling phase.
The source sends a \texttt{PrepForTransfer()} RPC to the target asynchronously,
transitioning the target to its own Target-Prepare phase.
The Target-Prepare phase tells the target that ownership transfer is
imminent. The target temporarily pends requests in the migrating hash
ranges (since some clients may discover the new views) and
services them after the source indicates that it has stopped servicing
requests in the old view.

\item[Transfer:]
Initiated after all source threads have completed the Prepare phase.
The source moves into its new view and stops servicing requests on the
migrating hash ranges.
When all server threads are in the new view, it sends out a
\texttt{TransferedOwnership()} RPC to the target asynchronously, which also
includes the hot records sampled in the Sampling phase.
This moves the target into its Target-Receive phase, whereupon it inserts the sampled
records into its \faster instance and then begins servicing requests for
the migrating hash ranges.
This also triggers the target to service any requests pending from the
Target-Prepare phase.

\item[Migrate:]
Initiated after all source threads have completed the Transfer phase.
The source uses thread-local sessions to send records in the migrating hash ranges
to the target.
Threads interleave processing normal requests with sending batches of migrating
records collected from the source's hash table to the target.
Each thread works on independent, non-overlapping hash table regions, avoiding
contention.

\item[Complete:]
Initiated after all source threads have completed the Migrate phase.
The source sends a \texttt{CompleteMigration()} RPC asynchronously, moving the target to the Target-Complete phase.
Then,
% the source asynchronously checkpoints its log, so if it crashes
% after migration has completed, it can be recovered without filtering out
% migrated records.
%
% When the checkpoint completes,
the source sets a flag in the metadata store
indicating that its role in migration is complete, and it returns to
normal operation.

\end{description}

\iffalse
\begin{table}[t]
\small
\begin{tabular}[]{p{0.95\columnwidth}}
\toprule
\textbf{Scale Out Phases On The Source Server} \\
\midrule
\textbf{Sample} \\
  \hspace{1em} $\diamond$ Copy records to the tail of the log on first access \\
  \\
\textbf{Prepare} \\
  \hspace{1em} $\diamond$ Prepare Target for ownership transfer \\
  \hspace{2em} $\rightarrow$ \texttt{PrepForTransfer()} RPC to Target \\
  \hspace{1em} $\diamond$ Complete all pending requests \\
  \\
\textbf{Transfer} \\
  \hspace{1em} $\diamond$ Change view \\
  \hspace{1em} $\diamond$ Transfer ownership and sampled records to the Target \\
  \hspace{2em} $\rightarrow$ \texttt{TransferOwnership()} RPC to Target \\
  \\
\textbf{Migrate} \\
  \hspace{1em} $\diamond$ Move records to the Target in parallel \\
  \hspace{2em} $\rightarrow$ \texttt{PushHashes()} RPCs to Target \\
  \\
\textbf{Complete} \\
  \hspace{1em} $\diamond$ Inform Target that migration has completed \\
  \hspace{2em} $\rightarrow$ \texttt{CompleteMigration()} RPC to Target \\
  \hspace{1em} $\diamond$ Take a checkpoint \\
  \hspace{1em} $\diamond$ Set vote at the Metadata store \\
  \\
\bottomrule
\end{tabular}
\caption{}
\label{table:source}
\end{table}
\fi

\begin{figure}[t]
\centering
\includegraphics[width=\columnwidth]{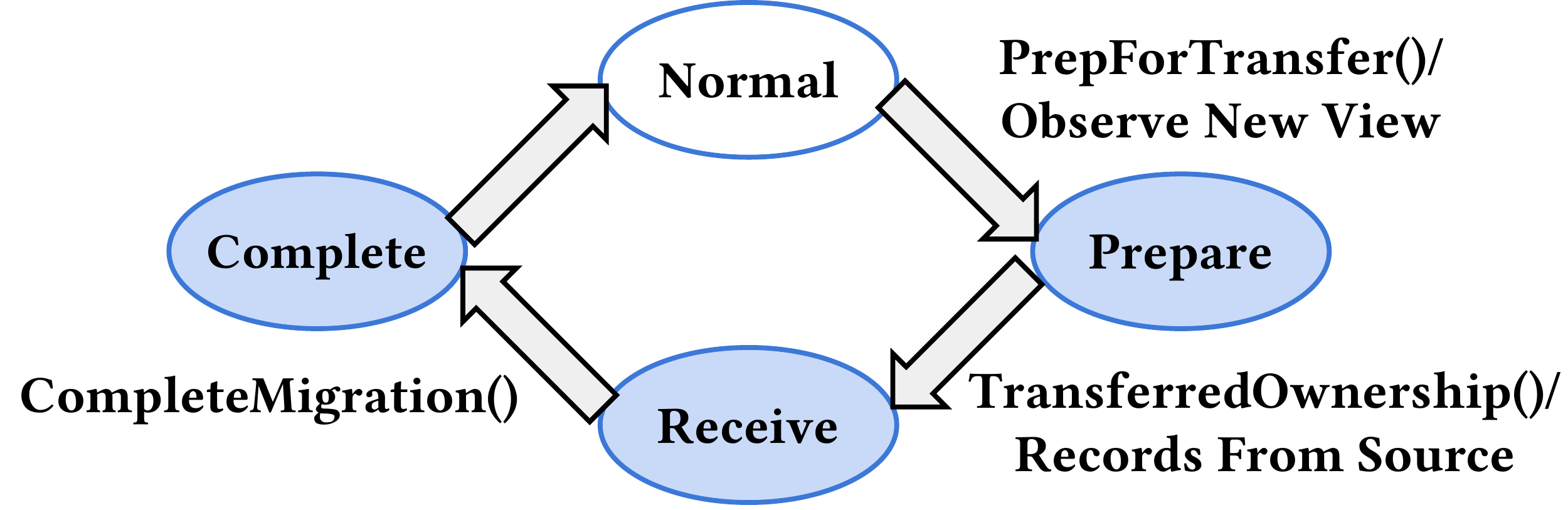}
\caption{Migration state machine on the target.}
%    It is
%    responsible for moving the target into the new view,
%    safely executing requests on the migrating hash
%    range, and receiving records from the source}
\label{fig:target}
\end{figure}

The target is mostly passive during migration; most of its phase changes are
triggered by source RPCs (Figure~\ref{fig:target}).
Requests for a record may arrive after a \texttt{TransferredOwnership()} RPC is received by the target, but before the
source has sent that record.
The target marks these requests pending, and it processes them when it receives
the corresponding record.

When the target receives the \texttt{CompleteMigration()} RPC,
% it also
% takes an asynchronous checkpoint, ensuring that it can be recovered to its
% post-migration state in the case that the source or target crashes.
%
% Afterward,
it also sets a flag at the metadata store indicating that its role in the
migration is complete, and it returns to normal operation.

Migration has succeeded once both servers have set their respective
flags at the metadata store.
A cluster management thread will have to periodically check these flags;
on finding both set, it deletes the
dependency at the metadata store to complete migration.

Shadowfax maintains high throughput during
scale up via low-coordination, non-blocking
epoch actions and purely asynchronous inter-machine communication.
The source prioritizes request processing, making progress in between request batches.
Its state machine transitions are independent of the target; all
migration RPCs and checkpoints are asynchronous.
The target prioritizes request processing in the same way.
Early ownership transfer, sampled records, and pending operations
let the target start servicing requests on moved ranges
quickly, improving throughput recovery.
Sessions let the source collect and
asynchronously transmit records in parallel while the target receives
and inserts them in parallel.

% Finally, the protocol is also wait-free; if either the source or target
% fail to make progress through it, an independent entity can always set a
% cancellation bit at the metadata store's migration dependency and roll
% back both servers.

\subsection{Leveraging Shared Storage for Decoupling}
\label{sec:design:indirection}

Migration cannot complete until all records have been
moved to the target, so Shadowfax must ensure that this
happens quickly.
However, \faster's larger-than-memory index makes this challenging:
entries in its hash table point to linked lists of records, which can
span onto local SSD.
Performing I/O (sequential or random) to migrate these records
can slow migration and hurt throughput.

Shadowfax's shared remote tier helps solve this problem.
Records on local SSD are always eventually flushed to this tier, so migration
can avoid accessing them.
When the source encounters an address for a record in a list that is on the
SSD, it sends an \emph{indirection record} to the target that indicates this
record's location in the shared tier.
This indirection record contains the next address in the list, an identifier for the
source's log, the hash range being migrated, and the hash entry that
pointed to the list.
The target inserts these records into its hash table using the hash
entry contained in the record.
%
% These records become part of its end-of-migration checkpoint.
%
Overall, these fine-grained inter-log dependencies represented by indirection
records accelerate migration completion by eliminating all I/O
that would otherwise be needed to consolidate records and transmit them
to the target.

During normal operation, if the target encounters an indirection record
when processing a request and the request's key falls in the
hash range contained in the record, the target asynchronously retrieves
the actual record from the
shared tier using the contained address and log identifier, inserts it
into its hash table, and then completes the request.

\subsection{Cleaning Up Indirection Records}

\begin{figure}[t]
\centering
\includegraphics[width=\columnwidth]{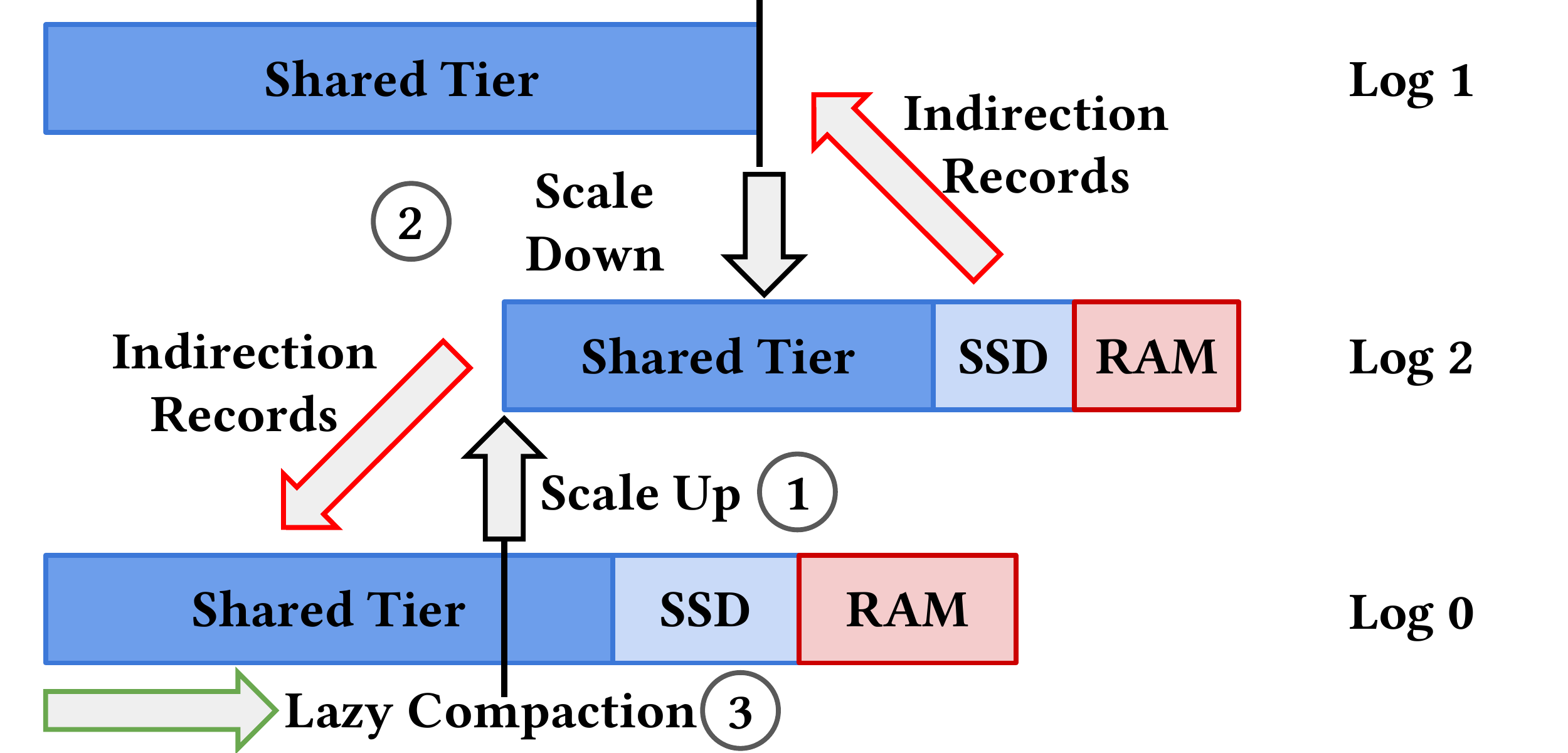}
\caption{Indirection records create inter-log dependencies.
%    fine-grained data dependencies
%    between logs.
%    These dependencies are cleaned up lazily during
%    log compaction.
%    This allows migration to be restricted to
%    main memory.
}
\label{fig:dfs}
\end{figure}

Migrations can accumulate indirection records between server logs
for records that are never accessed (Figure~\ref{fig:dfs}).
On scaling up (\one) by moving a hash range from
Log~0 to Log~2, Log~2 contains indirection records that point to
Log~0 on the shared tier.
Dependencies are also created during scale down (\two) when
records on Log~1 are migrated to Log~2.
These dependencies must eventually be cleaned up.

Shadowfax must already periodically do log compaction to eliminate stale
versions of records from its shared tier; resolving and removing indirection
records can be piggybacked on this process to eliminate overheads for
cleaning them (\three).
When compacting its log, if a server encounters a record belonging to a
hash range it no longer owns, the server transmits the record to the
current owner.
On receiving such a record, the owner first looks up the key.
If it encounters an indirection record while doing so and the key falls
in the contained hash range, then it means that the key was not retrieved
from the shared tier after migration.
In this case, the server inserts the received record; otherwise, it discards
the record.

Barring normal case request processing, this lazy approach
ensures that records not in main memory are accessed only once, during
the sequential I/O of compaction, which has to be done anyway.
It is also deadlock-free:
two servers might have indirection records pointing to each others' log,
but the resulting dependencies are cleaned up independently.

\subsection{Fault Tolerance}
\label{sec:fault-tolerance}

Migrations in Shadowfax can be easily made fault tolerant.
%
% In addition to marking their part of the migration completed at the
% metadata store
During their
respective \texttt{Complete} phases in the protocol,
the source and target would first have to take a checkpoint before
setting their flags at the metadata store.
This would make the migration durable;
% , and the temporary dependency
% between the source and target at the metadata store is garbage collected.
%
if either machine crashes hereafter, it can be independently recovered
from a checkpoint containing the effects of the migration.

If either server crashes during the process, recovery must involve both, which is why the
metadata store tracks the dependency between them.
This is because of early ownership transfer; during migration, the
target services operations on the migrating ranges, but many records belonging
to it may still be on the source.
When recovering a server, if Shadowfax finds a migration dependency
involving the server without both completion flags set, it cancels the
migration by setting a cancellation flag in the metadata server.
Then, it transfers ownership of hash ranges back to the source
(incrementing the source and target's view), restores both machines using
their pre-migration checkpoints, and recovers requests on hash
ranges that were issued during migration at the source.
This cancellation procedure ensures that migration is deadlock-free by effectively wrapping
the entire migration in a form of two-phase commit that supports unilateral abort~\cite{gray-notes}.
Migration need not lock or pause operation on the hash ranges under migration
except from the time that \texttt{TransferredOwnership} is issued until the
time that it is received.

%
%If either server fails to make progress through the protocol
%in a timely manner, the migration can always be cancelled by any party, and
%both servers can be rolled back.
%
%No server can stall migration completion indefinitely.
%
%\new{Overall, cancellation effectively wraps the entire migration in a form of
%two-phase commit that supports unilateral abort~\cite{gray-notes} where
%migration need not lock or pause operation on the hash ranges under migration
%except from the time that \texttt{TransferredOwnership} is issued until the
%time that it is received.}

\new{Another challenge with crashes
%that is orthogonal to migration
is in revocation of hash range ownership from an unavailable server to ensure it
does not accept requests in a stale view for hash ranges it no longer owns.
Views only help here if unanimity can be reached both among clients and
servers, which generally is not practical at scale. To solve this, Shadowfax can 
rely on classic lease-based approaches~\cite{leases,vertical-paxos}.}

We are working on implementing such crash recovery extensions as future 
work. For example, our recent work on \emph{distributed prefix 
recoverability}~\cite{dpr-sigmod21} addresses the problem of consistently recovering 
client sessions that span accesses to multiple shards.

%A full description and evaluation of these 
%mechanisms is beyond the scope of
%this paper, and we leave it to future work.

%
%As with ownership transfer, it creates a global cut across client
%sessions, indicating which operations clients should retransmit
%to servers after a crash.
%
%This client-assisted recovery eliminates the need for write-ahead-logging at
%servers, which would introduce a serial bottleneck.
%

\new{
\section{DISCUSSION}
\label{sec:discussion}

Shadowfax's techniques are not restricted to KVSs and can be applied to
other systems as well.
Its partitioned sessions can be used by \emph{stateful} cloud services
to preserve throughput over the network.
In fact, our implementation of sessions is templated on the service; we
used \faster for the purpose of this paper, but one could also use
parameter servers, graph stores, model serving systems etc.

Likewise, asynchronous global cuts can be used to scale out these
services while preserving throughput.
Since these cuts help propagate changes in ownership
across cores and machines, they can also be used for other operations
that involve changes in ownership like
failure detection and crash recovery.

Shadowfax's migration protocol can also be
used for scale in.
Since this protocol is fast and has low impact, it can also be used to
correctly partition records across servers.
In a distributed setting, partitioning becomes critical to
performance; pre-partitioning records between
servers results in load imbalances, which significantly hurts
throughput~\cite{dynamo, slicer}.
Migration allows Shadowfax to
dynamically partition its hash space into arbitrary, fine-grained splits
and
avoid pre-partitioning.
Using load information available at runtime, it can first determine the ideal
way to split its hash space across servers.
It can then quickly migrate these splits
between them.
View validation helps too; a server can own many fine-grained splits and still
serve 100~Mops/s.
}

\section{EVALUATION}
\label{sec:eval}

To evaluate Shadowfax, we focused on six key questions:
\begin{description}[leftmargin=0.5em]
\item[Does it preserve \faster{}'s performance?]
  \S\ref{sec:eval:clients} shows that Shadowfax preserves \faster{}'s
  scalability and adds in negligible overhead.
  Its throughput scales to 130~Mops/s on 64~threads on a VM even when using
  Linux TCP.

\item[How does it compare to an alternate design?]
  \S\ref{sec:eval:clients} shows that Shadowfax performs 4x better than
  a state-of-the-art approach that partitions dispatch as well as data.

\item[Does it provide low latency?]
  \S\ref{sec:eval:latency} shows that while serving a throughput of 130~Mops/s,
  Shadowfax's median latency is 1.3~ms on Linux TCP.
  Using two-sided RDMA decreases this to 40~\us.

\item[Can it maintain high throughput during scale out?]
  In \S\ref{sec:eval:migration}, we see that when migrating 10\% of a server's
  hash range, Shadowfax's scale-out protocol can maintain throughput
  above 80~Mops/s.
  Parallel data migration can help complete scale out in under 17~s,
  and sampled records help recover throughput 30\% faster
  (\S\ref{sec:eval:sampling}).

% \item[Can sharding be lazy and flexible?]
%   \S\ref{sec:eval:migration:views} shows that view validation has lower
%   runtime overhead compared to hash validation, which can deteriorate
%   throughput by upto 17\%.
%
%   When coupled with fast migration, this allows Shadowfax to be lazy and
%   flexible about how it distributes load.

\item[Do indirection records help scale out?]
  \S\ref{sec:eval:migration:indir} shows
  that by restricting migration to main memory, \new{indirection records avoid
  the cost of immediate post-migration I/O that other
  approaches require.}
  They also have a negligible impact on server throughput once scale out
  completes.
% , but at the cost of larger migrations.
%
% They have a negligible impact on normal case throughput, but result
% in larger pending queues.
%
% They can be cleaned up during compaction with no additional overhead.

\item[Do views reduce scale out's impact on normal operation?]
  In \S\ref{sec:eval:migration:views}, we show that validating ownership
  using views has a negligible impact on normal case server throughput.
  When compared to hash validating each request within a batch, views
  improve throughput by as much as 17\% depending on the number of hash
  ranges owned by the server.

\item[Can it scale across scales?]
  \S\ref{sec:eval:system-scalability} shows that when scaled across
  machines, Shadowfax continues to retain \faster's high throughput.
  \new{A cluster consisting of 768 threads spread across 12 servers scales
  linearly to 930~Mops/s while servicing 2304 client sessions.}

\end{description}

% We also ran Shadowfax on a small
% CloudLab~\cite{cloudlab} cluster consisting of 256 cores spread across 8
% servers
% and found that its throughput scales linearly to 400~Mops/s.
%
% We omit this experiment from this paper because of a lack of space.

\subsection{Experimental Setup}

\begin{table}[t]
\centering
\small
\begin{tabular}{p{0.15\columnwidth} p{0.75\columnwidth}}
\toprule
\textbf{CPU} & Xeon E5-2673~v4~2.3 GHz, 64 vCPUs in total
\\
\midrule
\textbf{RAM} & 432~GB
\\
\midrule
\textbf{SSD} & 96,000 IOPS, 500 MB/s sequential writes
\\
\midrule
\textbf{Network} & 30~Gbps, Hardware accelerated
\\
\midrule
\textbf{OS} & Ubuntu 18.04, Linux 5.0.0-1036-azure
\\
\bottomrule
\end{tabular}
\caption{Virtual machine details used to evaluate Shadowfax.}
%  This is the
%  E64\_v3 series available on Azure. Instances were configured to use
%  hardware accelerated networking.}
\label{table:exptconfig}
\end{table}

We evaluated Shadowfax on the Azure public cloud~\cite{azure}.
We ran all experiments on the E64\_v3 series of virtual
machines~\cite{e64} (Table~\ref{table:exptconfig}).
Experiments use 64 cores unless otherwise noted.
Each VM uses accelerated networking, which offloads
much of the networking stack onto FPGAs~\cite{accel-nw}, allowing us to
evaluate Shadowfax over regular Linux TCP.
Shadowfax's remote tier uses Azure's paged blobs on premium
storage~\cite{page-blobs}, which offer 7,500 random IOPS with a
throughput of 250~MB/s per blob.

We used a dataset of 250~million records, each consisting of
an 8~byte key and 256~byte value (totalling 80~GB in Shadowfax).
To evaluate the system under heavy ingest, we used YCSB's F
workload~\cite{ycsb} consisting of read-modify-write requests.
Each request reads a record, increments a counter within the record, and
writes back the result.
This counter could represent heartbeats for a sensor device, click
counts for an advertisement or views/likes on a social media profile.
Unless noted, requested keys follow YCSB's default Zipfian distribution ($\theta = 0.99$).
\new{
The experiments do not use checkpointing, which is needed for durability and to
bound recovery times. \faster{}'s checkpointing and durability scheme is described in
related work~\cite{cpr,dpr-sigmod21}.
}

We compare to two baselines; one representing the
state-of-the-art in fast request processing, the other representing the
state-of-the-art in data migration.

\begin{figure*}[t]
\centering
\begin{minipage}[t]{0.33\textwidth}
\includegraphics[width=\textwidth]{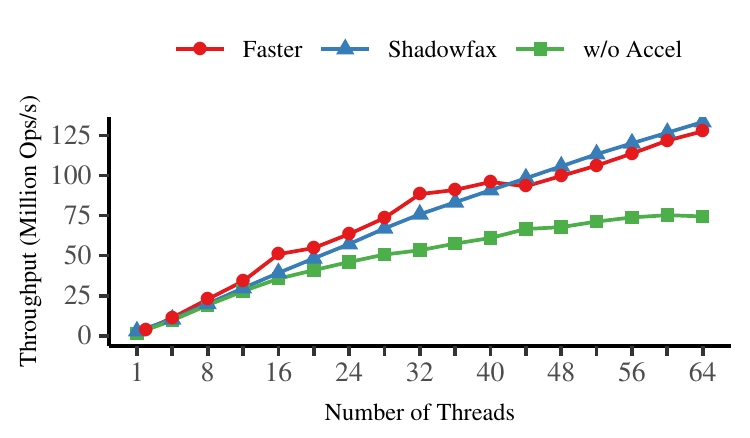}
\caption{Throughput scalability.}
%\caption{Shadowfax's thread scalability. With TCP acceleration enabled,
%    throughput scales linearly to 130 Mops/s and tracks \faster (with no networking).
%    With acceleration disabled, throughput scales to only 75 Mops/s.}
\label{fig:thread-scalability}
\end{minipage}%
\begin{minipage}[t]{0.33\textwidth}
\includegraphics[width=\textwidth]{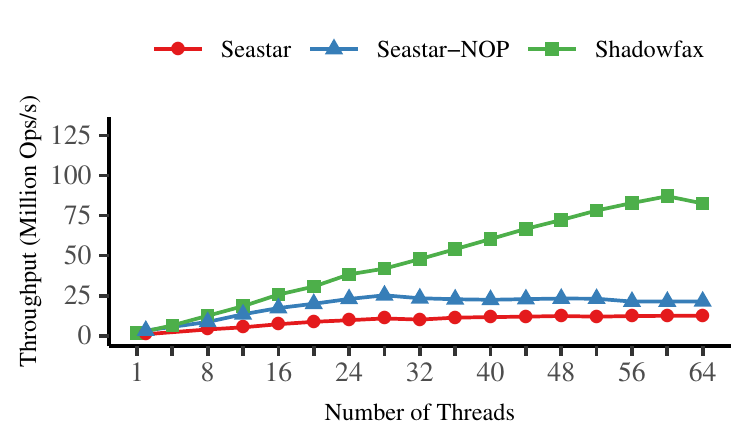}
\caption{\new{Shadowfax vs Seastar.}}
%\caption{With TCP acceleration,
%    throughput scales linearly to 87 Mops/s under a uniform distribution.
%    In comparison, Seastar scales to 10 Mops/sec.}
\label{fig:thread-scalability-comparison}
\end{minipage}%
\begin{minipage}[t]{0.33\textwidth}
\includegraphics[width=\textwidth]{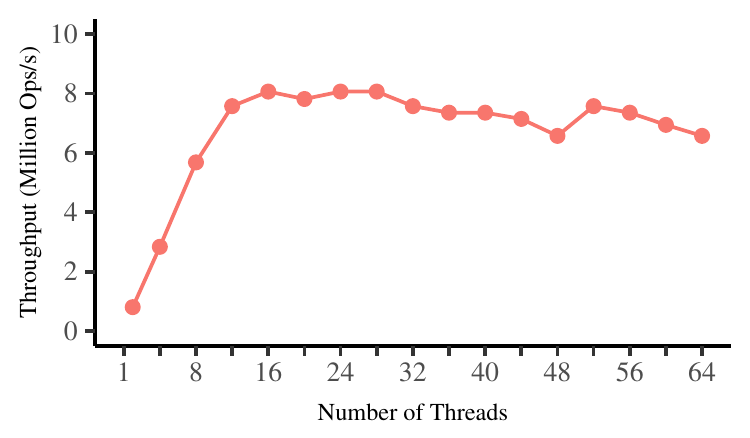}
\caption{\new{Insert-only workload.}}
%\caption{\new{For an insert-only workload, Shadowfax's throughput is
%    limited by the rate at which \faster's log tail can be incremented.
%    Throughput scales to 8~Mops/s on 16 threads and then saturates.}}
\label{fig:thread-scalability-append}
\end{minipage}
\iffalse
\caption*{Shadowfax's thread scalability. With TCP acceleration enabled,
    throughput scales linearly to 130 Mops/s and tracks \faster (with no networking).
    With acceleration disabled, throughput scales to only 75 Mops/s.
%
    Under a uniform distribution, Shadowfax is 8.5$\times{}$ better
    than Seastar, which scales to only 10 Mops/sec.
%
    \new{For an insert-only workload, Shadowfax's throughput is
    limited by the rate at which \faster's log tail can be incremented.}
}
\fi
\end{figure*}

\noindent
\textbf{Seastar+Memcached}~\cite{seastar}
is an open-source framework for building high
performance multi-core services.
Its shared-nothing design constrasts with Shadowfax;
servers partition data across cores, eliminating the need for locking.
Clients can send requests to any server thread;
Seastar uses message passing via shared memory queues to route
each request to the core that processes requests for that data item.
Seastar represents a best case for the state-of-practice; it is highly optimized.
It uses lightweight, asynchronous futures to avoid
context switch overheads, and it uses advanced NIC features like
FlowDirector~\cite{flow-director} to partition and scale network processing.
We used an open-source, lock-free, shared-nothing version of
Memcache on Seastar as a baseline~\cite{seastar-apps}.
We batched 100 operations per request, which maximized its throughput.

\noindent
\textbf{Rocksteady}~\cite{rocksteady} is a state-of-the-art migration
protocol for RAMCloud~\cite{ramcloud}.
To accelerate migration, it immediately routes requests for migrated records
to the target, while it is transfering records (which only reside in memory).
It slowly performs disk I/O in the background to incorporate the migrated records
into durable, on-disk replicas that belong to the target; this must complete
before the source and target can be independently recovered.
%
%A temporary, coarse-grained dependency between the source and
%target's on-disk data ensures fault tolerance if either machine
%crashes.
%
We modified Shadowfax to use a similar approach as a baseline.
Instead of using indirection records, first, all in-memory records are moved;
then, the source performs a sequential scan over all records on durable
storage, where all encountered live are sent to the target.
%
%Like in Rocksteady, the source and target stay coupled together for
%fault-tolerance during this phase.
%
%Once it completes, migration is completed, and they can be decoupled
%using Shadowfax's 2PC based mechanism.

\subsection{Throughput Scalability}
\label{sec:eval:clients}

Shadowfax partitions request dispatching across threads for
performance.
It shares access to \faster between threads to provide high
throughput even under skew.
To demonstrate this, we measured throughput while scaling the number of threads
on one server machine with one client machine.
The entire dataset resides in memory, ensuring the experiment is CPU-limited.
Figure~\ref{fig:thread-scalability} shows the results on Shadowfax, on
\faster when requests are generated on the same machine (i.e., no networking involved), and on Shadowfax without
hardware accelerated networking.

%{\color{red} RS: I think we need to be a bit careful here; is there a way we
%can word this paragraph still to reinforce that we aren't partitioning data or network
%load in any particular way?}
%
Shadowfax retains \faster{}'s scalability.
\faster{} scales to service 128~Mops/s on 64 threads.
Adding in the dispatch layer and remote client preserves performance;
Shadowfax scales to 130~Mops/s on 64 threads.
This is because it avoids cross-thread synchronization or communication for
request processing from the point a client thread issues a request until the
server thread executes it on \faster.
Client threads' pipelined batches of asynchronous requests also avoid any
slowdown from stalls induced by network delay, keeping all threads at the
client and server busy at all times.

Hardware network acceleration also plays an important role in
maintaining performance; when disabled, throughput reaches only 58\%
(75~Mops/s) of
accelerated TCP.
Here, CPU overhead for TCP transport processing increases, so the server
slows due to additional time spent in \texttt{recv()} syscalls instead of doing
work.
Hardware acceleration offloads a significant portion of packet processing to a
SmartNIC, allowing Shadowfax to maintain \faster{}'s scalability without
relying on kernel-bypass networking (DPDK or RDMA).

Next, we compared Shadowfax to Seastar+memcached
(Figure~\ref{fig:thread-scalability-comparison}) using a uniform key access
distribution; this is the only distribution that Seastar's client harness
supports (this advantages Seastar's shared-nothing approach, which suffers
imbalance under skew).
Seastar scales to 10~Mops/s on 28 threads, after which throughput is flat.
Shadowfax scales linearly to 85~Mops/s on 64 threads; even at 28 threads, it is
already 4x faster than Seastar.
This is because Seastar partitions work at the wrong layer; threads maintain
independent indices to avoid synchronizing on records, but this forces threads
to use inter-core message passing when they receive a request to route it to
the thread that has that record.
\new{
To ensure that this is the case and that it is not the result of a bottleneck in
Seastar's shared-nothing memcached implementation, we also measured the
throughput of Seastar's when each request is a no-op (by disabling its index,
see Seastar-NOP).
This improves Seastar's throughput, but it is still 4$\times{}$ slower
than Shadowfax on 64 threads.
This reinforces that simply attaching a more scalable index like
\faster to Seastar's networking and dispatch layers is not sufficient to get
good performance; forced cross-core routing of requests is the bottleneck.
}

In contrast, Shadowfax's design helps it exploit its shared \faster instance,
which is lock-free and minimizes cache footprint.
It leaves all synchronization and communication to the hardware
cache coherence, which is more efficient than explicit software
coordination and only incurs high costs when real contention arises in data
access patterns, rather than pessimistically synchronizing on all requests.
Shadowfax's advantage grows with skew;
\new{
comparing Figures~\ref{fig:thread-scalability}
and~\ref{fig:thread-scalability-comparison}
}
shows Shadowfax's performance
improves by 1.5x
under skew, whereas Seastar's performance would decrease.
%
%Seastar's client harness does not support generating a Zipfian
%distributed workload, but partitioning data records leads to load
%imbalance across threads under high skew, and hurts throughput.
%
%Because it avoids such imbalances and uses a highly optimized index,
%Shadowfax's throughput improves by 1.5x on moving from a uniform to a
%Zipfian distribution.

\new{
\subsubsection{Insert only workload}
\label{sec:eval:append-only}

\faster's \hlog is key to Shadowfax's high throughput since it allows
records to be updated in place.
However, in-place updates might not always be possible.
For workloads that are insert only, throughput will be limited by the
rate at which records can be appended to the \hlog's tail.
Figure~\ref{fig:thread-scalability-append} presents scalability for a
workload that inserts 250~million records into Shadowfax.
Throughput scales to 8~Mops/s on 16 threads.
Beyond 16 threads, increments to the \hlog's tail bottleneck the system,
and throughput saturates.
}

\subsection{Batching and Latency}
\label{sec:eval:latency}

Shadowfax clients send requests in pipelined batches to amortize
network overheads and keep servers busy.
Asynchronous requests with hardware network acceleration
help reduce batch sizes and latency.
To show this, we measured its median latency and batch size at
server saturation.
Table~\ref{table:latency} shows results with TCP, TCP with
hardware acceleration disabled, and two-sided RDMA (Infrc).
We used Azure's HC44rs~\cite{hc44} instances for Infrc, since they support
(100~Gbps) RDMA; they have Xeon Platinum 8168s with 44 vCPUs.

Most of Shadowfax's latency comes from batching, which amortizes CPU
costs. Accelerated networking reduces CPU load, decreasing the batching
needed to retain throughput.
%
%The asynchronous client library prevents threads from blocking while
%waiting for these in-flight batches to complete at the server, and
%allows them to continue issuing requests into session buffers.
%
%Accelerated networking keeps the batch size required to saturate
With acceleration, small 32~KB batches saturate server throughput with a low
latency low of 1.3~ms.
%
%With hardware acceleration disabled, the software network stack now
%becomes a bottleneck (\texttt{recv()}).
%
Without acceleration, increased batch size doesn't help; with 32~KB
batches throughput drops to 75~Mops/s, and median latency increases to 2.2~ms.
\new{Finally, the TCP~1~KB case uses a small batch size with hardware
acceleration; latency drops by 6.1$\times$ but throughput also drops by
6.8$\times$ showing the combined importance of acceleration and proper batch
size.}

\begin{table}[t]
\centering
\small
\begin{tabular}{p{0.18\columnwidth}
                >{\centering\arraybackslash}m{0.20\columnwidth}
                >{\centering\arraybackslash}m{0.11\columnwidth}
                >{\centering\arraybackslash}m{0.16\columnwidth}
                >{\centering\arraybackslash}m{0.09\columnwidth}}
\toprule
\textbf{Network} & \textbf{Throughput (Mops/s)} &
\textbf{Batching (KB)} & \textbf{Median Latency (\us)} &
\textbf{Queue Depth}
\\
\midrule
\textbf{TCP} & 130 & 32 & 1300 & 1927
\\
\midrule
\textbf{\new{TCP, 1 KB}} & \new{19} & \new{1} & \new{212} & \new{60}
\\
\midrule
\textbf{w/o Accel} & 75 & 32 & 2200 & 1927
\\
\midrule
\textbf{Infrc} & 126 & 1 & 38.6 & 60
\\
\midrule
\textbf{TCP-IPoIB} & 125 & 8 & 260 & 482
\\
\bottomrule
\end{tabular}
\caption{\new{Shadowfax's latency at server saturation.}}
% On Azure's RDMA
% instances, it can maintain a median of
% 40~\us while performing 126~Mops/s. With TCP,
% this increases to 1.3~ms.}
\label{table:latency}
\end{table}

\begin{figure}[t]
\centering
\includegraphics[width=0.75\columnwidth]{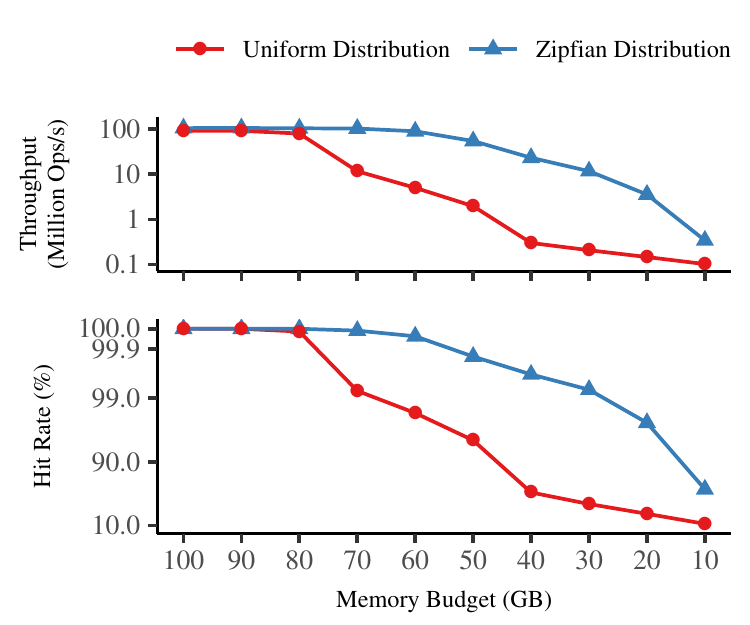}
\caption{\new{Throughput under decreasing memory budgets.
%    Under a
%    Zipfian access pattern, it can sustain high throughput under
%    small budgets because of a small working set that fits in memory.
}}
\label{fig:memory}
\end{figure}

\begin{figure*}[t]
\centering
\includegraphics[width=1.0\textwidth]{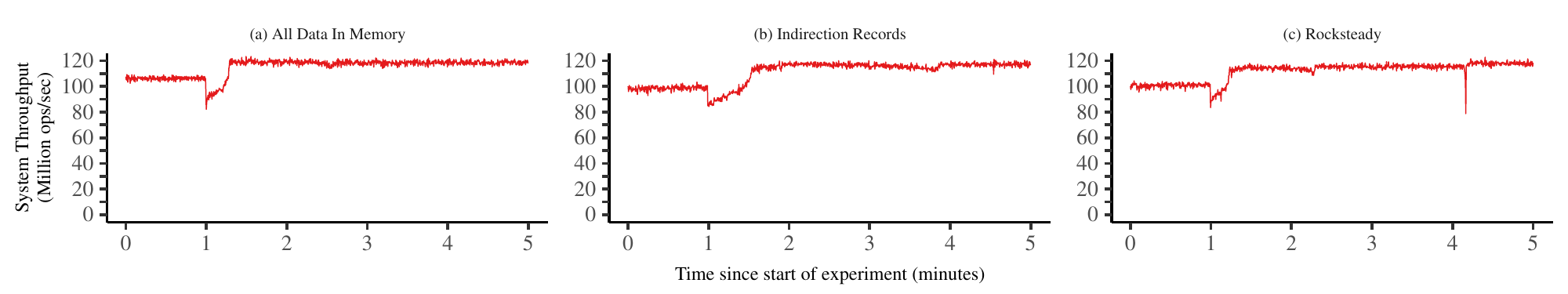}
\caption{Running throughput when 10\% of a server's load is migrated
    to an idle target.}
%    Migration was initiated at 1 minute. For a
%    memory budget of 60~GB (graph (b)),
%    scale-out shifts load in 32~s while maintaining
%    throughput above 80 Mops/sec.}
\label{fig:migration}
\end{figure*}

\begin{figure*}[t]
\centering
\includegraphics[width=1.0\textwidth]{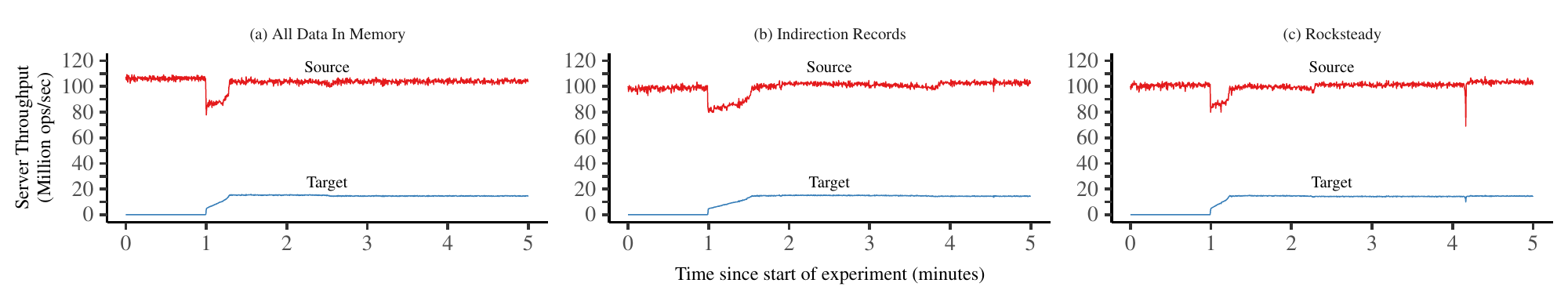}
\caption{Source and target throughput during scale up.}
%    Sequentially
%    scanning over the cold tier during migration (graph (c)), increases
%    the duration of scale out to 180~s during which the source
%    loses one thread's worth of throughput (1.5~Mops/sec).}
\label{fig:migration-split}
\end{figure*}
%
% These instances are able to saturate Shadowfax at 44 vCPUs
% itself.
%
% This is because Shadowfax is compute bound under our workload, and
% cores on the Infrc instances can simultaneously turbo boost to 3.7~GHz
% compared to 2.6~GHz on the TCP instances.
%
% As a result, they can saturate throughput with fewer vCPUs.

The batch size required to saturate throughput on Infrc is
significantly lower at 1~KB, dropping median latency to 40~\us.
This is because the network is faster and the stack is implemented in hardware;
servers and clients can receive and transmit batches with near-zero
software overhead (including system calls).
Secondly, vCPUs on these instances are faster with a base clock rate of
2.7~GHz compared to 2.3~GHz on the TCP instances
(Table~\ref{table:exptconfig}).
This speeds servers and clients, reducing the batch size and
threads (from 64 to 44) required to reach the same throughput.
To evaluate this further, we ran Shadowfax using TCP over IPoIB~\cite{ipoib} on
the Infrc instances (Table~\ref{table:latency}, TCP-IPoIB).
Throughput still saturates at 125~Mops/s.
Compared to hardware accelerated TCP, faster vCPUs reduce the batch size
by 4x (8~KB) and median latency by 5x (260~\us).
%
%Differences in the network might also contribute to these improvements,
%but we found Shadowfax to be CPU-bound in both cases.

%This demonstrates an important benefit of Shadowfax's design.
%
%Instead of improving performance by increasing
%client fan-in at a server, the system can be sped up by
%simply improving performance at each thread; either by using faster
%vCPUs (cores) or by optimizing software.

\new{
\subsection{Memory Budget}
\label{sec:eval:dfs}

% \begin{figure}[t]
% \centering
% \includegraphics[width=\columnwidth]{
%     sofaster-data/memory-budget/memory-budget-disk.pdf}
% \caption{}
% \label{}
% \end{figure}

\faster's throughput eventually becomes limited by the SSD when the
entire dataset does not fit in main memory.
Shadowfax's dispatch layer and client library ensure
that this does not change when requests are generated over the cloud
network.
To show this, we measured throughput under a decreasing main-memory
budget for the \hlog.
We also measured the hit rate (the percentage of requests
that were served from main-memory) during this experiment.
Figure~\ref{fig:memory} presents the results
(please note the log scale).

Overall, throughput drops as the memory budget decreases.
This is because the system needs to issue random I/O to fetch records
from SSD.
Once fetched, these records are appended to the \hlog's tail which
flushes records at its head to SSD leading to more random I/O during
future requests.
For a uniform distribution, throughput begins to drop at
80~GB.
Since all records are equally hot, even a small set on SSD hurts the hit
rate and saturates SSD IOPS (Table~\ref{table:exptconfig}).
For a Zipfian distribution, a smaller hot set ensures that this
begins to happen only at 50~GB.
Throughput still drops because of low SSD IOPS
(Azure throttled our VMs to 96,000 IOPS), decreasing to 3.5~Mops/s at 20~GB.
However, this is still 24$\times{}$ better than the uniform case which
drops to 0.146~Mops/s.

}

\subsection{Scale Out}
\label{sec:eval:migration}

\begin{figure*}[t]
\centering
\includegraphics[width=1.0\textwidth]{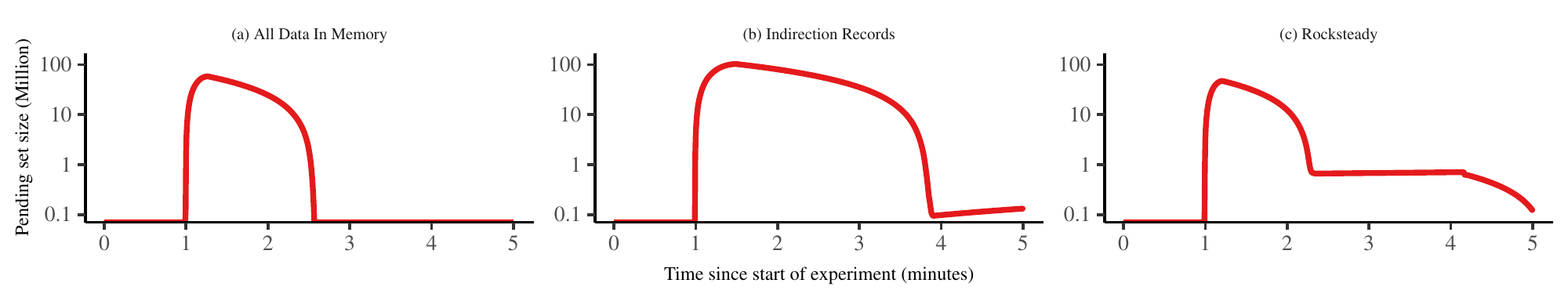}
\caption{Number of pending operations during scale up.}
%    Indirection
%    records (graph (b)) result in remote accesses to shared storage,
%    which leads to larger pending queues once scale up has completed.
%    Without them (graphs (a), (c)), these queues drain shortly after
%    scale up completes.}
\label{fig:comparison-pending}
\end{figure*}

Shadowfax's migration transfers hash ranges between
two machines and minimizes throughput impact while doing so.
Indirection records help restrict migration to memory, speeding up
scale out, decoupling the source and target sooner.
To demonstrate this, we measured throughput during scale up.

In a 5-minute experiment with one client and two servers (a source and a
target), the entire hash space initially resides at the source.
After one minute, 10\% of this hash range is moved to the target.
Figure~\ref{fig:migration} shows system throughput during
the experiment; Figure~\ref{fig:migration-split}
shows source and target throughput separately.
In (a), all records are placed in memory.
In (b) and (c), servers are restricted to a memory budget of 60~GB,
allowing us to compare the impact of indirection records (in (b)) against
Rocksteady's scan-the-log approach (in (c)).

\subsubsection{All-In-Memory Scale Out}

Global cuts for ownership transfer avoid stalling cores at migration start, but
the view change for this cut has some impact; request batches are invalidated,
causing requests to be shuffled among sessions buffers at the client
\new{
($\approx$250,000 requests per view change based on Table~\ref{table:latency} Queue Depths).
}
This is visible in Figure~\ref{fig:migration} (a); throughput at the
start of scale out (1 minute) briefly drops to 80~Mops/s.

Figure~\ref{fig:migration-split} (a) shows that throughput on the source
stays at 85~Mops/s after this.
This is because the source is collecting and transmitting records as it
services requests.
Parallel migration limits the length of this impact in two ways.
First, it accelerates migration, completing in 17~s and restoring full
throughput.
Second, as more records shift to the target, it
serves more requests, causing system throughput to recover
even before scale up completes.
Once scale up completes, system throughput increases by 10\% as expected.

Shadowfax's asynchronous client library helps limit the
impact too.
When the target receives a request for a record that has not been
migrated yet, it marks the request as pending.
This keeps clients from blocking, allowing them to
continue sending requests.
To prevent a buildup of pending requests, the target periodically tries to complete them.
Figure~\ref{fig:comparison-pending} (a) shows the number of pending operations at the
target during migration.
When migration starts, requests flood the target, pending 100 million requests.
As records migrate, these requests complete, with the last pending operation
completing 100~s after migration start.
Hence, practical migrations must be small and incremental to bound
delay; however, throughput recovery is more important in Shadowfax's target
applications whereas latency can be tolerated with asynchrony.

\new{
We also ran the above experiment on a larger cluster of four 64-core machines (3 servers, 1 client)
on CloudLab~\cite{cloudlab} and obtained similar results; aggregate cluster
throughput is only impacted by 20\% in the worst case
during migration, since throughput is only reduced at the source during
migration.
}

\subsubsection{Indirection Records}
\label{sec:eval:migration:indir}

With a 60~GB memory budget, some records to be migrated are on the source's
SSD.
Rocksteady's approach (Figure~\ref{fig:migration} (c)) migrates records from
memory and then scans the on-SSD log to migrate colder records.
Parallel migration completes the in-memory phase in just 14~s.
Thoughput improves quickly after this phase, since these are hotter records.
\new{
However, the second phase is single threaded, scans over files on SSD, and
takes 165~s to complete; during this phase the source and target remain
inter-dependent for fault tolerance.}

\begin{table}[t]
\centering
\small
\begin{tabular}{p{0.35\columnwidth} p{0.55\columnwidth}}
\toprule
\textbf{Config} & \textbf{Data Migrated (GB)}
\\
\midrule
\textbf{All Data In Memory} & 7.44
\\
\midrule
\textbf{Indirection Records} & 16.47
\\
\midrule
\textbf{Rocksteady} & 5.60
\\
\bottomrule
\end{tabular}
\caption{Impact of indirection records on migration size.}
%    Indirection
%    records lead to larger migrations because records that are not in
%    main memory cannot be filtered out.}
\label{table:migration-size}
\end{table}

% \begin{figure}[t]
% \centering
% \includegraphics[width=0.9\columnwidth]{graphs/migration-size.pdf}
% \end{figure}

\begin{figure*}[t]
\centering
\begin{minipage}[t]{0.33\textwidth}
\includegraphics[width=\textwidth]{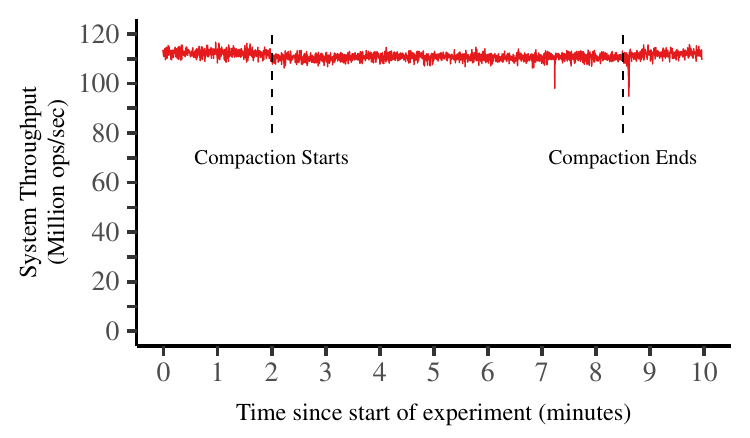}
\caption{Cleaning indirection records.}
%\caption{Running throughput when compacting the source's log and
%    cleaning up indirection records while doing so. Indirection records
%    do not add any additional overhead over that of compaction.}
\label{fig:compaction}
\end{minipage}%
\begin{minipage}[t]{0.33\textwidth}
\includegraphics[width=\textwidth]{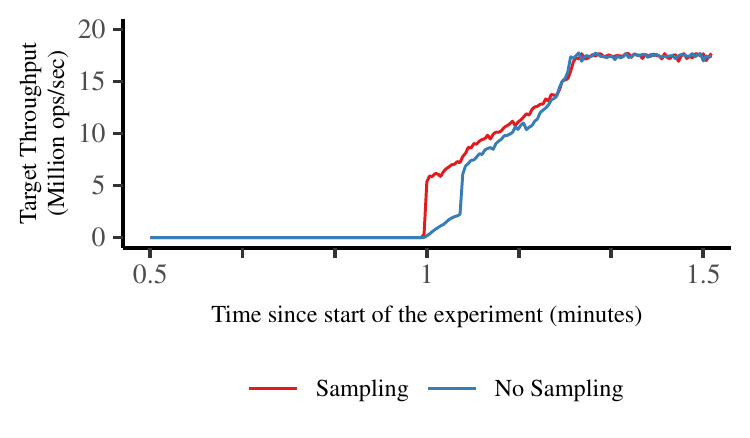}
\caption{Sampled records impact.}
%\caption{Impact of shipping sampled records during ownership transfer to
%    the target. Sending them over improves the target's
%    throughput during the first 5~s of migration.}
\label{fig:migration-sampling}
\end{minipage}%
\begin{minipage}[t]{0.33\textwidth}
\includegraphics[width=\textwidth]{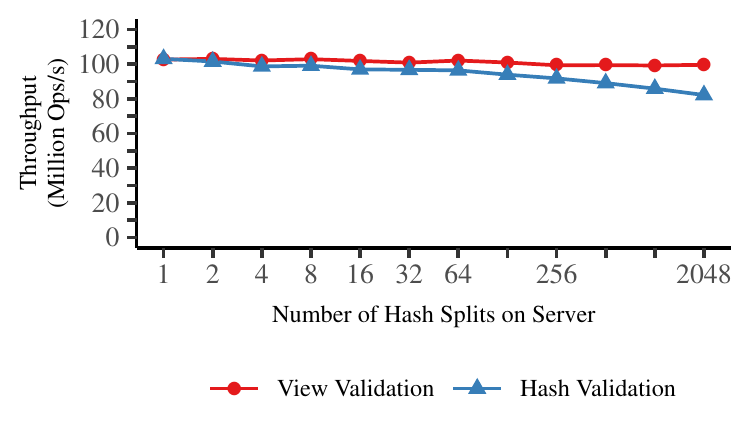}
\caption{View validation overhead.}
%\caption{The overhead of using views to validate record ownership
%    at a server is negligible. When coupled with fast migration, this
%    allows Shadowfax to shard and redistribute load whenever required.}
\label{fig:migration-views}
\end{minipage}
\iffalse
\caption*{Migration optimizations.
%
    Indirection records
    do not add any additional overhead over that of compaction; running
    throughput remains the same when compacting the source's log and
    cleaning up indirection records while doing so.
%
    Sending over sampled records improves the target's
    throughput during the first 5~s of migration.
%
    The overhead of using views to validate record ownership
    at a server is negligible. This
    allows Shadowfax to shard and redistribute load whenever required.
}
\fi
\end{figure*}

\new{
Indirection records solve this, completing migration in 32~s
(Figure~\ref{fig:migration} (b)) by avoiding this I/O as part
of migration.}
By sending out records that point to shared remote storage, migration
is restricted to memory and avoids I/O at the source altogether.
However, this approach increases the amount of data transmitted to the target.
Table~\ref{table:migration-size} show this effect.
Compared to Rocksteady's 5.60~GB, indirection records cause
16.47~GB to be transmitted from memory to the target.
This is because we must send about one indirection record per hash table bucket
entry, totaling 11~GB here.
%
%We leave improving this algorithm to future work.
%
The larger migration takes 18~s longer than Rocksteady's in-memory phase, but it
decreases the total duration of migration by 150~s.

After migration, requests that hit indirection records at the
target cause remote accesses to shared cloud storage.
These requests are infrequent (these records are
cold), and they have little impact on throughput
(Figure~\ref{fig:migration} (b)).
However, cloud storage is slow, so in the time it takes to retrieve one such
record, the target receives many requests for it which must pend.
Requests that pend during scale out complete by 4 minutes
(Figure~\ref{fig:comparison-pending} (b)).
The gradual upward slope after this is due to the requests that
pend on access to remote shared storage.
Requests never pend after scale out with Rocksteady; however, its slow
sequential scan causes requests to pend awaiting transmission from the source
during its longer migration.

We also measured the impact of fetching records from shared remote storage when
resolving indirection records during compaction, but its throughput impact was
neglible (Figure~\ref{fig:compaction}).

\subsubsection{Sampled Records}
\label{sec:eval:sampling}

Shadowfax sends a small set of hot records to the target during ownership
transfer, which allows the target to start servicing requests and recovering
throughput quickly.
Figure~\ref{fig:migration-sampling} shows target throughput
when this is enabled (Sampling) and when it is disabled (No
Sampling).
In this experiment, all data starts in the source's memory, so scale out
completes in 17~s.
When enabled, throughput at the target rises up to 8~Mops/s immediately
after ownership transfer.
If disabled, this happens 5~s later, once sufficient records have
been migrated over.
At this point, nearly 30\% of scale out has completed, meaning that
by sampling and shipping hot records during ownership transfer,
the target starts contributing to system throughput 30\% faster.
\new{
Measurements on the source show that the \texttt{SAMPLING} phase lasted
4~ms and had no noticeable overhead.
}

\subsubsection{Ownership Validation}
\label{sec:eval:migration:views}

Views allow Shadowfax to fluidly move ownership of hash ranges between
servers and help minimize the overhead of scale out on normal
operation of the system.
Figure~\ref{fig:migration-views} demonstrates this; it presents normal
case server throughput under an increasing number of hash splits.
When using views to validate record ownership at the
server (View Validation), throughput stays fairly constant.
On switching over to an approach that hashes every received key and
looks up a trie of owned hash ranges at the server (Hash Validation),
throughput gradually drops as the number of hash splits increase.

This figure shows the benefit of using views given a particular scale
out granularity; if scale out always moves 7\% of a server's load (16
hash splits), then view validation can improve normal case
throughput by 5\%.
Similarly, if it always moves 0.2\% of a server's load (512 hash
splits), then this improvement increases to 10\%.

\subsection{System Scalability}
\label{sec:eval:system-scalability}
\begin{figure}[t]
\centering
\includegraphics[width=0.9\columnwidth]{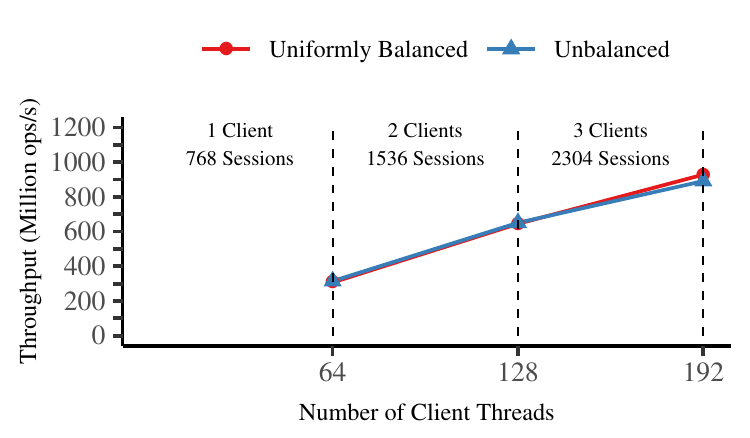}
\caption{\new{Shadowfax system scalability.}
%    Shadowfax continues to retain FASTER’s
%    high throughput even across servers. A cluster consisting of
%    12 servers scales linearly to 930 Mops/s.}
%
%     while
%     servicing 1024 client sessions.
}
\label{fig:system-scalability}
\end{figure}

% Shadowfax's session's layer allows client threads to directly connect
% with server threads.
%
In addition to retaining \faster's throughput within a machine,
Shadowfax also retains throughput across machines.
\new{
To demonstrate this, we first hash partitioned 2 billion records across a
cluster consisting of 12 servers on
CloudLab~\cite{cloudlab} (each server had 64 threads, 128 GB RAM and
one 100~Gbps Mellanox CX5 NIC).
}
Next, we measured the total throughput of this cluster while varying the
number
of clients issuing requests (clients had the same hardware as servers).
\new{
Because each client thread opens up a session to one thread on each server,
each client added in 64 sessions to each server and hence 768 sessions
to the cluster (64 threads/client * 12 servers).
}

Figure~\ref{fig:system-scalability} shows the results.
\new{
"Unbalanced" shows results for a Zipfian skewed workload.
Cluster throughput scales to 890~Mops/s but
sub-linearly when moving from two
clients to three.
}
This was with 12 coarse-grained hash ranges, one per server.
This is insufficient to uniformly distribute load across servers.
%
% Shadowfax
% would work well with and benefit from techniques that partition and
% distribute at a finer granularity and replicate hot keys~\cite{dynamo, slicer}.
%
% In fact, because these techniques tend to frequently rebalance load, its
% scale-out protocol would be critical to them.
%
\new{
Shadowfax's migration is designed to fix this
via fine-grained hash splits.
Load distributions can be monitored to
determine ideal hash splits~\cite{slicer}.
Once determined, these splits can be quickly migrated with low
throughput impact.
"Uniformly Balanced" (Figure~\ref{fig:system-scalability}) shows
an upper bound that could be achieved this way.
It represents a case where splits uniformly distribute load
over all servers,
improving throughput by 40~Mops/s (4.5\%) to 930~Mops/s.
}

Finally, beyond high throughput, this experiment also demonstrates that
Shadowfax can scale to support a large number of client sessions
(connections); at
saturation, each server has 192 sessions open to it, resulting in a
total of 2304 sessions across the cluster.

\section{RELATED WORK}
\label{sec:related}

Shadowfax builds on several areas of recent
research.

\noindent
{\bf Epochs and Cuts.}
There are many schemes for synchronization and memory protection in lock-free
concurrent data structures including hazard pointers~\cite{hazard},
read-copy-update~\cite{rcu} and epoch-based schemes~\cite{epochs, epochs-phd}.
Like \faster{} and Shadowfax, several other
systems~\cite{bwtree,deuteronomy,deuteronomy-ranges,hekaton-indexing} use epochs for this purpose.
%
% Notably, many of these systems combine epoch protection with copy-on-write or
% log-structured updates, which adds overhead due to allocation and tail-of-log
% contention. Other systems use update-in-place with latching~\cite{masstree} to avoid this overhead, but target main memory. Shadowfax
% uses update-in-place for values in memory, and switches to copy-on-write via a lock-free
% hash index when values are stable; its unique approach to epochs is key to avoiding overhead
% when determining which values require copy-on-write.
%
Shadowfax's use of epochs to avoid strong ordering among requests except on coarse 
boundaries resembles Silo's, a (single-node) in-memory store~\cite{silo}.
%Like in Silo, Shadowfax's epochs avoid strong
%ordering among requests except on coarse boundaries, improving scalability.
%Silo also uses epochs to improve write-ahead logging
%scalability~\cite{silor}.
Shadowfax extends epochs back to clients by
asynchronously choosing points in server execution and correlating these back
to per-client sequence numbers, effectively pushing the overhead of logging out
of servers altogether. Similarly, Scalog's persistence-before-ordering approach uses
global cuts that define and order shards of operations on different
machines~\cite{scalog}.
%; Shadowfax uses similar cuts across threads and client
%session buffers to define an order to enforce boundaries among operations.

\noindent
{\bf High-throughput Networked Stores.}
Some in-memory stores exploit kernel-bypass networking or RDMA and optimize
for multicore. Many of these focus on throughput but do not provide
scale out~\cite{pilaf,mica,herd}, both of which
can slow normal-case request processing. RAMCloud focuses on low latency
and has migration, but its throughput is two
orders of magnitude less than Shadowfax~\cite{ramcloud,ramcloud-recovery}.
%
% Anna~\cite{Wu:2019:ATC:3311880.3322438} uses selective key replication with
% a per-thread shared nothing design, but relies on eventual consistency with
% lazy gossip.
%
FaRM~\cite{farm,farm-tx}
%creates large clusters of distributed memory where
%clients primarily
uses one-sided RDMA reads to construct data structures like
hash tables and supports scale out via in-memory replication.
%FaRM supports scale out and crash recovery 
%by relying on
%whole-machine battery backup and in-memory replication.
FaRM's reported per-core throughput is about 300,000 reads/s/core, compared to Shadowfax's
1.5~million read-modify-writes/s/core, though there are differences in
experimental set up. For example, FaRM doesn't report numbers for
read-modify-write or write-only workloads which are significantly more
expensive in FaRM, since they involve server CPU, require replication, and
cannot be done with one-sided RDMA operations.

% Elasticity

\noindent
{\bf Elasticity.}
Scale out and migration are key features in shared,
replicated stores~\cite{dynamo,cassandra,redis}.  High-throughput, multicore
stores complicate this because normal-case request processing is highly
optimized and migration competes for CPU.  Some stores rely on
in-memory replicas for fast load redistribution~\cite{farm-tx,drtm-migration};
this is expensive due to DRAM's high cost and
replication overhead.
% , so this does not work for Shadowfax.
%
Squall~\cite{squall} migrates data in the H-Store~\cite{hstore}
database; it exploits skew via on-demand record pulls from
source to target with colder data moved in the background.
Rocksteady~\cite{rocksteady} uses this idea in RAMCloud along with a
deferred replication scheme that avoids write-ahead logging.
% for migrated data.

\section{CONCLUSION}
\label{sec:conclude}

Practical KVSs must ingest events over the network and elastically scale across machines.
Shadowfax does this with state-of-the-art performance that
reaches 130~Mops/s/VM by relying on its global cuts, partitioned
sessions, and end-to-end asynchronous clients.

\section*{ACKNOWLEDGMENTS}
 This work was started at Microsoft Research during an internship by Chinmay Kulkarni and a visit by Ryan Stutsman. We thank Donald Kossmann and the anonymous reviewers for their 
 comments and suggestions.
 Chinmay Kulkarni is supported by a Google PhD Fellowship, which partially
 supported this work.
 This material is based upon work supported by the National Science
 Foundation under Grant No.\ CNS-1750558. Any opinions,
 findings, and conclusions or recommendations expressed in this material are
 those of the author(s) and do not necessarily reflect the views of the
 National Science Foundation.

{\footnotesize
\bibliographystyle{acm}
\bibliography{bib}
}

\end{document}